\documentclass[]{aa}  
\AtBeginDocument{\nolinenumbers}   % turn the numbering off for the whole document
\let\linenumbers\relax              % stop the class from turning it back on

\usepackage{graphicx}
\usepackage{txfonts}
\usepackage{lipsum}
\usepackage{subcaption}         % necessary for continued figures, example in section 3
\usepackage{lscape}             % to rotate a single page table, example in appendix.
\usepackage{placeins}           % useful with \FloatBarrier, to keep 
\usepackage[                    % For coloured references
  colorlinks=true,              
  linkcolor=blue,               % internal links (sections, refs)
  citecolor=blue,               % citations
  urlcolor=blue                 % URLs
]{hyperref}
\usepackage{siunitx}
\providecommand{\unit}{\si}

\newcommand{\lya}{{\text{Ly}\ensuremath{\alpha}}}
\newcommand{\Halpha}{{\text{H}\ensuremath{\alpha}}}
\newcommand{\Hbeta}{{\text{H}\ensuremath{\beta}}}
\newcommand{\Hgamma}{{\text{H}\ensuremath{\gamma}}}
\newcommand{\fesclya}{\ensuremath{f_{\rm esc}^{\lya}}}
\newcommand{\fesclyc}{\ensuremath{f_{\rm esc}^{\rm LyC}}}

\newcommand{\Hden}{\ensuremath{n_{\rm H}}}
\newcommand{\NH}{\ensuremath{N_{\rm H}}}

\newcommand{\eden}{\ensuremath{n_{\rm e}}}

\newcommand{\logten}{\ensuremath{\log_{10}}}

\newcommand{\logU}{\ensuremath{\logten U}}
\newcommand{\logNO}{\ensuremath{\logten {\rm (N/O)}}}
\newcommand{\logCO}{\ensuremath{\logten {\rm (C/O)}}}
\newcommand{\logNC}{\ensuremath{\logten {\rm (N/C)}}}
\newcommand{\Msun}{\ensuremath{M_\odot}}
\newcommand{\Zsun}{\ensuremath{Z_\odot}}

\newcommand{\MUV}{\ensuremath{M_{\rm UV}}}
\newcommand{\xiion}{\ensuremath{\xi_{\rm ion}}}

\begin{document}

%%%%%%%%%%%%%%%%%%%%%%%%%%%%%%%%%%%%%%%%
% if you use custom commands in your title,
% ensure to check your title when submitting!
%%%%%%%%%%%%%%%%%%%%%%%%%%%%%%%%%%%%%%%%
   \title{A Little Red Dot emerging from a nitrogen-rich UV-luminous galaxy with Very Massive Stars}
   \titlerunning{COLA1: An LRD emerging from a nitrogen-rich UV-luminous galaxy with VMSs}

   \author{Alberto~Torralba\inst{\ref{inst:ista}}\corrauth{alberto.torralba@ista.ac.at}
   \and Jorryt~Matthee\inst{\ref{inst:ista}}
   \and Rui~Marques-Chaves\inst{\ref{inst:geneva_obs}}
   \and Fabrice~Martins\inst{\ref{inst:LUPM}}
   \and Benjamín~Navarrete\inst{\ref{inst:ista}}
   \and Gabriele~Pezzulli\inst{\ref{inst:kapteyn}}
   \and Edoardo~Iani\inst{\ref{inst:ista}}
   \and Rohan~P.~Naidu\inst{\ref{inst:hawaii}}
   \and Gabriel~Brammer\inst{\ref{inst:dawn}, \ref{inst:nbi}}
   \and Claudia~Di~Cesare\inst{\ref{inst:ista}}
   \and Anne~Hutter\inst{\ref{inst:uniwien}}
   \and Gauri~Kotiwale\inst{\ref{inst:ista}}
   \and Ruari~Mackenzie\inst{\ref{inst:EPFL}}
   \and Sara~Mascia\inst{\ref{inst:ista}}
   \and Romain~Meyer\inst{\ref{inst:geneva_obs}}
   \and Daniel~Schaerer\inst{\ref{inst:geneva_obs}}
   }

   \institute{
   Institute of Science and Technology Austria (ISTA), Am Campus 1, 3400, Klosterneuburg, Austria.\label{inst:ista}
   \and Geneva Observatory, Department of Astronomy, University of Geneva, Chemin Pegasi 51, 1290, Versoix, Switzerland.\label{inst:geneva_obs}
   \and LUPM, CNRS \& Univ. Montpellier, Place Eugene Bataillon, {34095}, {Montpellier}, {France}\label{inst:LUPM}
   \and Kapteyn Astronomical Institute, University of Groningen, {Landleven 12}, {9747}, {Groningen}, {Netherlands}\label{inst:kapteyn}
   \and Institute for Astronomy, University of Hawai‘i, {2680 Woodlawn Drive}, {96822}, {Honolulu}, {USA}\label{inst:hawaii}
   \and Cosmic Dawn Center (DAWN), Copenhagen, Denmark\label{inst:dawn}
   \and Niels Bohr Institute, University of Copenhagen, Jagtvej 128, Copenhagen, Denmark\label{inst:nbi}
   \and Department of Astrophysics, University of Vienna, {T\"urkenschanzstr. 17}, {1180}, {Vienna}, {Austria}\label{inst:uniwien}
   \and Institute of Physics, Laboratory of Astrophysics, EPFL, Observatoire de Sauverny, {Chemin Pegasi 51}, {1290}, {Versoix}, {Switzerland}\label{inst:EPFL}
}

   \date{\today}

  \abstract{
  JWST is revealing abundant populations of high-$z$ galaxies with unforeseen properties: UV-luminous galaxies with high nitrogen enrichment, populations of very massive stars (VMSs; $>100\ M_\odot$), and early forms of massive black hole growth such as the Little Red Dots (LRDs). Here we present all these phenomena appearing simultaneously in COLA1, a luminous ($M_{\rm UV}=-21.3$) galaxy at $z=6.6$. New NIRSpec medium-resolution data reveal a blue UV continuum ($\beta_{\rm UV}=-2.5$) with strong emission lines including N\,{\sc iv}] $\lambda1486$, stellar P~Cygni profiles, and broad He\,{\sc ii} $\lambda1640$. We fit the UV with stellar population models, and find that the models including VMSs (limited to $300\ M_\odot$) are statistically preferred. The best-fit stellar population has $M_*\approx10^8\ M_\odot$, and a young age of $2.4\pm0.8$ Myr. The rest-optical shows a redder continuum ($\beta_{\rm opt}=-1.1$) and broad Balmer lines with Balmer absorption akin to LRD spectra. By evolving the best-fit stellar population we find that the UV could dim by 1--2 mag in $\sim5$ Myr, and hence this system would be observed as a typical V-shaped LRD. We measure a supersolar $\rm log_{10}(N/O)=-0.07\pm0.05$ and subsolar $\rm log_{10}(C/O)=-0.64\pm0.11$; therefore $\rm log_{10}(N/C)=+0.58\pm0.12$. The independent density-sensitive N\,{\sc iv}] $\lambda\lambda1483,1486$ and C\,{\sc iii}] $\lambda\lambda1906/1909$ ratios require two zones with $n_e\gtrsim10^{6.5}$~cm$^{-3}$ and $n_e\lesssim10^{3.5}$~cm$^{-3}$, respectively. We estimate $\rm log_{10}(N/C)>1.3$ (2$\sigma$) for the dense zone, requiring extreme N/C enrichment mechanisms such as highly rotating $>100\ M_\odot$ VMSs, tidal disruption events, or supermassive stars. We speculate that runaway collisions of massive stars in a dense stellar cluster may have lead to the seeding of an intermediate-mass black hole resulting in the formation of an LRD.
  }


   \maketitle

\section{Introduction}\label{sec:intro}

JWST has already delivered transformative results for our understanding of the earliest galaxies and black holes \citep[for a review, see e.g.,][]{Matthee2025_review, Ubler2026_review, Stark2026_review}. In particular, JWST has unveiled an abundant population of luminous galaxies ($\MUV\lesssim -20$) up to $z\approx 14$ \citep[e.g.,][]{Castellano2022, Naidu2022_z10, Harikane2023, Donnan2024, Whitler2025, Carniani2024, Naidu2026_momz14}, exceeding the expectations based on pre-JWST predictions \citep[e.g.,][]{Bouwens2015, Bouwens2021, Oesch2018}. What physical mechanisms are responsible for the prevalence of such UV-luminous galaxies at early times is a subject of debate. Some possible explanations include an increased star-formation efficiency in low-mass halos \citep[e.g.,][]{Inayoshi2022, Harikane2023, Dekel2023}; stochastic star-formation at early times, which introduces significant scatter biasing the bright end of the UV luminosity function \citep[e.g.,][]{Sun2023, Shen2023, Mirocha2023, Mason2023}; or the presence of exotic AGN and/or stellar populations with possibly top-heavy initial mass functions \citep[IMFs; e.g.,][]{Trinca2024, Yang2024, Harikane2023, Hutter2025}.

Deep spectroscopy of the most luminous galaxies at high redshift has revealed unexpected properties, such as strong UV nitrogen emission, suggesting that many UV luminous galaxies are highly N-enhanced \citep[e.g.,][]{Schaerer2024, Ji2024, Castellano2024, Naidu2026_momz14, Chen2026_gnz11, Nakane2026}. In standard chemical evolution models, moderate nitrogen enrichment occurs through core-collapse supernovae (CCSNe). However, significant N-enhancement can only happen in timescales of $\sim 100$~Myr, from the winds of asymptotic giant branch stars \citep[AGB; e.g.,][]{Vincenzo2016, Kobayashi2020}. The fact that some of the most N-enhanced systems are observed at only a few hundreds of Myr after the Big Bang demands alternative, faster routes of enrichment \citep[e.g.,][]{Isobe2023, Senchyna2024, Watanabe2024}. Some processes that could explain the observed chemical abundances include a high incidence of populations of Wolf-Rayet stars \citep[e.g.,][]{Berg2026, Kobayashi2024, Senchyna2024}, tidal disruption events (TDEs) in dense star clusters \citep[e.g.,][]{Isobe2023, Cameron2023, Watanabe2024}, winds of Very Massive Stars \citep[VMSs, $\gtrsim 100$~\Msun{}; e.g.,][]{Vink2023, Shi2026}, or supermassive stars \citep[SMS, $\gtrsim 10^3$~\Msun{}; e.g.,][]{Denissenkov2014, Charbonnel2023, Nagele2023}.
Very recent works have reported robust detection of VMSs populations in luminous high-$z$ galaxies \citep[][]{MarquesChaves2026_vms, Yang2026, Chen2026_gnz11, Nakane2026_gnz11}, identified through prominent stellar P~Cygni features and broad \ion{He}{ii} in the UV spectra \citep[e.g.,][]{Upadhyaya2024}. Some of these galaxies are also strong \ion{N}{iv}] emitters with high implied N/O enhancement, for example, CEERS-1019 \citep[][]{Larson2023} at $z=8.7$ or GN-z11 at $z=10.6$ \citep[][]{Oesch2016, Bunker2023}. Therefore, these galaxies could provide an observational link between nitrogen enhancement in early galaxies and their featured stellar populations.

Another surprising and extensively discussed JWST result is the discovery of an abundant population of red compact objects, the Little Red Dots \citep[LRDs; e.g.,][]{Labbe2023, Kocevski2023, Matthee2024, Greene2024}. The origin and nature of the LRDs' central engines is highly debated, with current hypothesis ranging from super-Eddington accreting supermassive black holes (SMBHs) obscured by nuclear dust \citep[e.g.,][]{Madau2026_se_lrds, Scholtz2026, Billand2026}, to intermediate-mass black holes ($M_{\rm BH}\sim 10^3$--$10^5$~\Msun{}; IMBHs) enshrouded in dense gas that reprocesses their radiation \citep[e.g.,][]{Ji2025_blackthunder, Naidu2025_BHStar, deGraaff2025, Liu2025, Naidu2026_etacar, Sneppen2026_sirocco, Kido2025}, to supermassive stars \citep[e.g.,][]{Nandal2026, Zwick2026, Chisholm2026, Kokorev2026, Lu2026}, or quasi-stars \citep[QS;][]{Begelman2026, Martins2026, Cantiello2025, Santarelli2026}. Regardless of the physical interpretation, the LRDs seem to represent a common and distinct phase of massive black hole growth at high-$z$, that becomes increasingly rare towards the nearby Universe \citep[e.g.,][]{Lin2026_local_1, Lin2026_desi, Park2026_desi, Kapoor2026, Loiacono2026}.

Some LRD systems seem to have very low metallicities, suggestive that they were formed in near-pristine environments \citep[e.g.,][]{Maiolino2026_pristine, Ivey2026}, although metal lines are often present in their spectra, such as \ion{Fe}{ii}, indicating some metal enrichment \citep[e.g.,][]{Lin2026_local_1, Torralba2026_gn9771, Deugenio2026_irony, Kokorev2026_glimpse}.
Remarkably, some luminous LRDs show prominent UV lines, including strong nitrogen emission, suggesting that the UV source of some LRDs could share the enrichment patterns of some of the most luminous high-$z$ galaxies \citep[e.g.,][]{Labbe2024, Isobe2025, Torralba2026_monster, Tang2026_spurs_NC, Morel2026, Isobe2025}.
Recent works have reported that the chemical abundances of LRDs and high-$z$ broad-line sources in general are compatible with globular clusters in formation \citep[e.g.,][]{Isobe2025, Ji2026_NO, Chisholm2026, Kokorev2026}.
Constraining the global metal content of LRDs (M/H) further allows to rule out formation pathways. For example, IMBHs with masses up to $\sim 10^5$~\Msun{} can form through direct collapse of near-pristine gas, if a nearby luminous source of UV photons is present that is capable of dissociating H$_2$, suppressing molecular cooling and preventing fragmentation \citep[e.g.,][]{Omukai2001, Oh2002}. For a massive black hole to be formed through direct collapse and rapid accretion, the metallicity of the gas must not exceed $\sim 10^{-3}$--$10^{-4}$~\Zsun{} \citep[e.g.,][]{Chon2020, Chon2026_sms_cluster}. This process could lead to the formation of an SMS as an intermediate step \citep[e.g.,][]{Bromm2003, Agarwal2012, Latif2013}; and/or a dense nuclear star cluster \citep[e.g.,][]{Omukai2008, Devecchi2009, Gieles2018} which would ultimately lead to ``seeding'' of a massive black hole. The direct-collapse scenario has received attention in the LRD context since many of these objects show nearby companions that could have acted as the UV sources needed to supress molecular cooling \citep[][]{Baggen2026, Barger2026, Pacucci2026}.

Here we present a spectroscopic analysis of COLA1 \citep[][]{Hu2016, Matthee2018}, a luminous galaxy at $z=6.6$ that we found to be a strong N-emitter, presents signs of containing VMSs, and shows broad lines in the optical spectrum \citep[][]{Meyer2026_cola1} similar to LRDs. The new JWST/NIRSpec data of this source unifies three of the most puzzling recent findings from JWST into a single source.
The structure of this paper is organized as follows. In Sect.~\ref{sec:data} we describe the observations and data reduction. In Sect.~\ref{sec:results} we present our empirical results, including stellar population fitting, emission line modeling and chemical abundance estimation. In Sect.~\ref{sec:discussion} we discuss physical interpretations of our results. Finally, in Sect.~\ref{sec:summary} we summarize our main findings.
Throughout this paper we adopt a $\Lambda$CDM cosmology as described by \citeauthor{Planck2020}~(\citeyear{Planck2020}; $\Omega_\Lambda = 0.69$, $\Omega_{\rm m} = 0.31$, $H_0 = 67.7$~\unit{km.s^{-1}.Mpc^{-1}}). All photometric magnitudes are given in the AB system \citep[][]{Oke1983}.

\section{Observations and data reduction}\label{sec:data}

\begin{figure*}
    \centering
    \includegraphics[width=\linewidth]{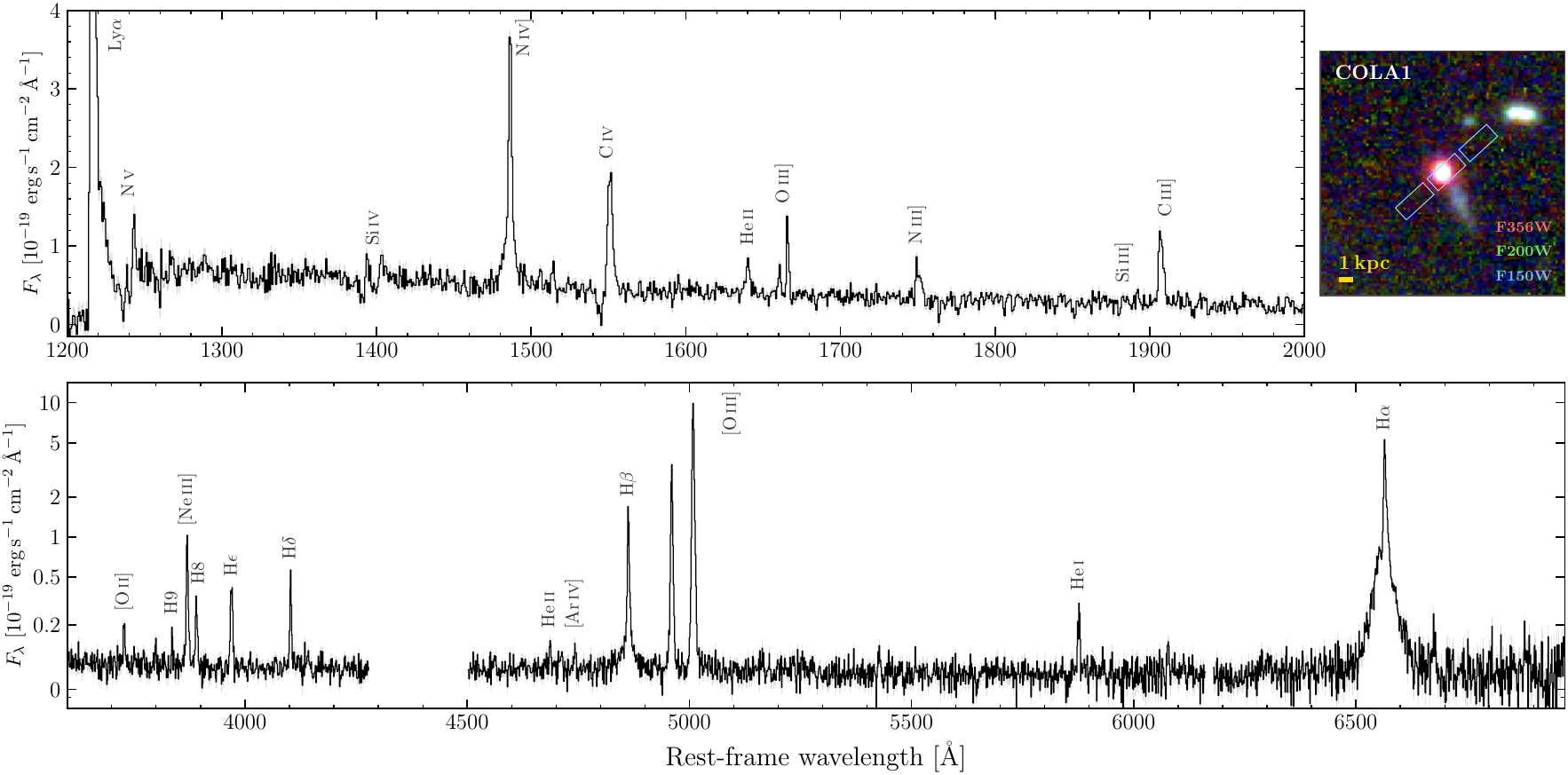}
    \caption{
    Full NIRSpec G140M+G235M rest-UV (top) and rest-optical (bottom) spectrum of COLA1. The UV is rich in strong emission lines, highlighting \ion{N}{iv}] $\lambda 1486$ that is suggestive of high N-enrichment. The UV also presents deep P~Cygni absorption profiles in \ion{N}{v}, \ion{Si}{iv} and \ion{C}{iv}; together with broad \ion{He}{ii} are indicators of a population of massive stars. The rest-optical shows strong a broad \Halpha{} and \Hbeta{}, with a superimposed absorption in \Halpha{} (see Fig.~\ref{fig:Ha_Hb_O3}), and strong narrow \ion{He}{i} $\lambda 5876$.
    }
    \label{fig:fig1}
\end{figure*}

\subsection{COLA1: A double-peaked \lya{} emitter at $z=6.6$}

COLA1 is a luminous galaxy ($\MUV = -21.3$) at $z=6.6$ \citep{Hu2016}, and an exceptionally strong Lyman-$\alpha$ (\lya{}) emitter with a double peaked profile. The mere observability of the blue \lya{} peak at such high redshift, when cosmic reionization is still incomplete \citep[e.g.,][]{Greig2017, Bosman2022}, implies that COLA1 is embedded in a large ionized region that allows enhanced transmission of \lya{} photons, redshifted by the Hubble expansion before they encounter a neutral intergalactic medium \citep[IGM; e.g.,][]{Matthee2018, Mason2020}.
The fact that COLA1 does not reside in a particularly large overdensity of galaxies \citep[][]{Torralba2024} indicates that it is a prolific producer and leaker of ionizing photons, and it should be capable of carving the ionized region virtually alone \citep{Meyer2026_cola1}, but also that we may be observing it along a rare, underdense line of sight \citep[][]{Mason2020, Gronke2021, Park2026_cola1}.
Some relevant general properties of COLA1 derived throughout this paper are listed in Table~\ref{tab:properties}.

\subsection{JWST/NIRSpec}

COLA1 was observed as primary target by the JWST Cycle 4 program \#7722 (PIs: Torralba \& Matthee), using NIRSpec/MOS. Spectroscopy was taken using the grating/filter combinations G140M/F070LP and a G235M/F170LP, with a total exposure time of 25\,209.6~s per grating ($\sim7$~h). A 3-point nodding pattern was used. The MSA data were reduced using the latest version of the software \textsc{msaexp}\footnote{\url{https://github.com/gbrammer/msaexp}} \citep[v0.9.17;][]{Heintz2024, deGraaff2025_rubies}, which builds upon the \texttt{jwst} pipeline\footnote{\url{https://github.com/spacetelescope/jwst}} (v1.16.1). We use the CRDS context file \texttt{jwst\_1464.pmap}. We use the extended-wavelength reduction, which extracts the spectra from the full coverage beyond the nominal range of the gratings, extrapolating the flat-field and photometric reference files so that the trace can be extracted across the full detector.
The flat-field was undone at the calibration stage and applied at extraction, and the empirical read-noise rescaling was turned off. The 1D spectra were extracted with optimal extraction \citep[][]{Horne1986} using a fitted profile, with the background obtained by differencing the nod positions and the trace position fixed from the MSA source position and nod offsets.
The MSA mask was designed so the trace of COLA1 does not overlap with that of any other object, and so COLA1 is placed in an optimal position on the detector to retrieve up to the observed \Halpha{} wavelength on G235M.
We show the spectra in Fig.~\ref{fig:fig1}, highlighting the most relevant features.

By combining the two gratings observed by the program GO \#7722 we obtain a complete coverage of the rest-UV-to-optical spectrum of COLA1; from $\sim 1000$ to $\sim 7000$~\AA{} rest wavelength. However, the detector gap falls on top of the critical \Hgamma{} and [\ion{O}{iii}] $\lambda 4363$ emission lines. We complement our analysis with the JWST NIRSpec/IFU data taken by GO \#3767 (PI: Meyer); 40~min on-source exposure time with the grating G395M. Those data provide a sufficient S/N of the \Hgamma{}+[\ion{O}{iii}] $\lambda 4363$ emission lines. We retrieved the IFU datacube from the DAWN JWST Archive\footnote{\url{https://dawn-cph.github.io/dja/index.html}}. The 1D spectrum was obtained using the optimal extraction method \citep{Horne1986}, with a 2D Gaussian extraction kernel, fitted to the morphology of COLA1 in the whitelight image of the IFU cube. the line fluxes of the overlapping MSA (G235M) and IFU (G395M) generally agree within uncertainties. For consistency, we rescale the flux of the IFU 1D spectrum to match the  [\ion{O}{iii}] $\lambda\lambda 4960,5008$ flux expected from slit-loss correction uncertainties.

\subsection{JWST/NIRCam imaging and morphological analysis}

We also use photometric images from JWST/NIRCam to measure the UV size of COLA1. In particular, we use imaging in the F115W, F150W, and F200W filters from the JWST GO program \#1933 \cite[PIs: Matthee \& Naidu; presented in][]{Torralba2024}. We fit the morphology of COLA1 with a Sérsic profile, and we use a second Sérsic to model the faint foreground contaminant \citep[see][]{Torralba2024}. We fit using the \texttt{pysersic} software \citep[][]{Pasha2023} and the synthetic PSFs from \texttt{stpsf} \citep[][]{Perrin2014}. We find effective radii of $R_{\rm eff} = 209 \pm 18$~pc and $R_{\rm eff} = 193 \pm 15$~pc in the F115W and F150W bands, respectively; the size is completely unresolved in F200W and F356W. Therefore, we adopt a fiducial UV size of $\approx 200$~pc. Note that this size is slightly smaller than the PSF half-radii ($\approx 230$~pc and $250$~pc for F115W and F150W, respectively), so COLA1 is only marginally resolved in the rest-frame UV.

\subsection{ALMA far-IR upper limits}

ALMA observations of COLA1 were conducted in Band 6 during Cycle 7 (project 2019.1.00741.S, PI: Matthee), using compact-to-intermediate 12m array configurations with 27--46 antennas across six execution blocks and a total on-source time of $\sim266$~min. The spectral setup consisted of two sidebands, each composed of two spectral windows with total bandwidths of 3.75 GHz and channel width resolutions of 7.8 MHz, and the main one aimed at the redshifted [\ion{C}{ii}] $\lambda 158$~\unit{\mu m} line. The observations have a synthesized beam of 0.374$^{\prime\prime}$. The data were calibrated using the automatic pipeline provided by the Common Astronomy Software Applications package \citep[CASA;][]{McMullin2007}, version 6.2.1.7.

From the non-detection in the integrated continuum band, we estimate a continuum upper $3\sigma$ limit of $<\,24.6$~\unit{\mu Jy.beam^{-1}} at observed $242$~\unit{GHz}.
The [\ion{C}{ii}] $\lambda 158$~$\mu$m is also undetected. We estimate an upper 3$\sigma$ limit of $L_{[\ion{C}{ii}]}<2\times 10^7\ L_\odot$, assuming a top-hat line with $\rm FWHM = 125$~\unit{km.s^{-1}}. This upper limit corresponds to an upper limit of $\logten(\rm SFR / \Msun{}\,yr^{-1})\lesssim 0.2$ using extrapolations of the standard calibrations \citep[e.g.,][]{DeLooze2014, Schaerer2020}, but see \citet{Harikane2018,Matthee2019} for discussions whether these could be applicable to LAEs. For comparison, the narrow \Halpha{} emission implies a somewhat higher $\logten(\rm SFR / \Msun{}\,yr^{-1})\sim 0.9$ using the calibrations from \citet[][]{Kramarenko2026}.

\section{Results}\label{sec:results}

\begin{table}
\centering
\caption{General properties of COLA1.}
\label{tab:properties}
\begin{tabular}{ccc}
\hline
Property & Value & Units \\
\hline
$M_{\rm UV}$ & $-21.36 \pm 0.01$ & -- \\
$\beta_{\rm UV}$ & $-2.5 \pm 0.1$ & -- \\
$\beta_{\rm opt}$ & $-1.1 \pm 0.7$ & -- \\
$r_{\rm eff, UV}$ & $0.21 \pm 0.02$ & kpc \\
$L_\Halpha$ (Total) & $9.9^{+0.8}_{0.3} \times 10^{42}$ & \unit{erg\,s^{-1}} \\
$L_\Halpha$ (Narrow) & $2.10^{+0.28}_{-0.14} \times 10^{42}$ & \unit{erg\,s^{-1}} \\
\fesclya & $0.67 \pm 0.3$ & -- \\
$\xi_{\rm ion} / (1 - \fesclyc)$ & $10^{25.56 \pm 0.02}$ & \unit{Hz.erg^{-1}} \\
$12 + \logten(\rm O/H)$ & $7.62 \pm 0.01$ & -- \\
\hline
\end{tabular}
\end{table}

\subsection{Flux measurements of UV and optical lines}\label{sec:emline_measurements}

\begin{figure}
    \centering
    \includegraphics[width=\linewidth]{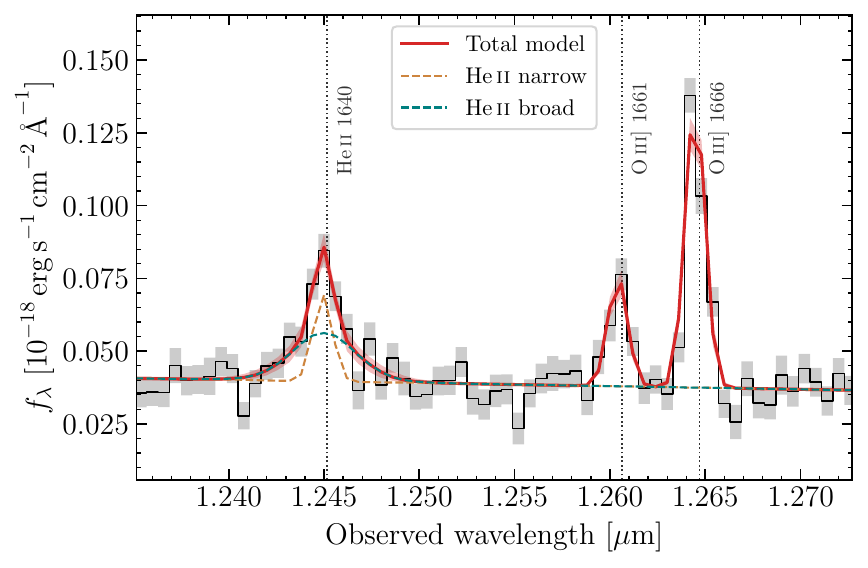}
    \caption{Fit to the \ion{He}{ii} $\lambda1640$ emission line and \ion{O}{iii}] $\lambda\lambda1600,1666$ doublet. The model fit is described in Sect.~\ref{sec:emline_measurements}. We use a broad + narrow Gaussian model for \ion{He}{ii} $\lambda1640$, and independent Gaussians for the \ion{O}{iii}] doublet.}
    \label{fig:fit_heii}
\end{figure}

The data shown in Fig.~\ref{fig:fig1} reveals strong UV emission lines including \ion{N}{iv}] $\lambda\lambda 1486,1483$, \ion{N}{iii}] $\lambda 1750$ complex, \ion{C}{iii}] $\lambda\lambda 1907,1909$,  broad \ion{He}{ii} $\lambda1640$ (Fig.~\ref{fig:fit_heii}), P~Cygni features in \ion{N}{v} $\lambda\lambda 1239,1243$, \ion{Si}{iv}] $\lambda\lambda 1394,1403$, and \ion{C}{iv} $\lambda\lambda 1548,1551$, and a very blue continuum slope ($\beta_{\rm UV} = -2.5\pm 0.1$). In the rest-optical, it shows broad Balmer lines, \Halpha{} absorption, and a redder continuum ($\beta_{\rm opt} = -1.1 \pm 0.7$).

We use the Python package \textsc{unite} \citep{Hviding2026_unite} to measure the emission lines of the G140M and G235M spectra of COLA1. We first performed a preliminary fit to the most luminous narrow emission lines in the rest-optical (i.e., [\ion{O}{iii}] $\lambda\lambda 4960,5008$, [\ion{Ne}{iii}] $\lambda 3869$, $\lambda3968$) to pinpoint COLA1's systemic [\ion{O}{iii}] redshift at $z=6.59054 \pm 0.00002$. Then, we fit every emission line as a single Gaussian, adding a second broad Gaussian component when needed.
The absorptions of the lines showing P~Cygni profiles are fitted as Gaussians with negative fluxes and blueshifted velocities with respect to the emission counterparts.
Broad components are added after a visual inspection, to fit the lines \ion{N}{iv}] $\lambda\lambda 1483, 1486$, \ion{C}{iv} $\lambda\lambda 1548, 1559$, \ion{He}{ii} $\lambda 1640$, [\ion{Ne}{iii}] $\lambda 3869$ and [\ion{O}{iii}] $\lambda\lambda 4960, 5008$. In addition, the \Halpha{} and \Hbeta{} emission lines present broad features, but these will be treated separately in Sect.~\ref{sec:balmer_lines_modeling}.
The full list of measured emission line fluxes is listed in Tables~\ref{tab:emlines_uv} and~\ref{tab:emlines_opt}. The fitted components for all the emission lines are shown in Figs.~\ref{fig:all_uv_lines} and \ref{fig:all_optical_lines}.

\subsection{Gas-phase O/H metallicity}\label{sec:metallicity}

From the emission line measurements we estimate the single-zone gas-phase metallicity of COLA1 using the direct method \citep[e.g.,][]{Peimbert1967}. We obtain a mean electron density of $\logten(\eden/\unit{cm^{-3}}) = 3.4^{+0.3}_{-0.4}$ from the \ion{[O}{ii}] $\lambda\lambda 3727,3729$ doublet. Another density tracer is the UV doublet \ion{C}{iii}] $\lambda\lambda1906,1909$, however the measured ratio is close to the theoretical maximum, so it lies in a regime where it is hardly constraining (see Sect.~\ref{sec:abundances}). We estimate a direct-method temperature of $(2.2 \pm 0.2) \times 10^4$~K from the [\ion{O}{iii}] $\lambda4363$ and [\ion{O}{iii}] $\lambda5008$ fluxes using PyNeb \citep[][]{Luridiana2015}. Finally, we estimated a gas-phase metallicity of $12 + \logten(\rm O/H) = 7.62 \pm 0.01$. From this result, we adopt a fiducial metallicity of $\approx 0.1$~\Zsun{} throughout this paper. However, note that this value should be interpreted with caution, given the evidence for  a multiphase ISM with heterogeneous enrichment within COLA1 (see Sect.~\ref{sec:abundances}). None of our key interpretations rely on a precise measurement of the oxygen abundance.

\subsection{Stellar population synthesis fitting}\label{sec:ficus_analysis}

\begin{figure}[t!]
    \centering
    \includegraphics[width=\linewidth]{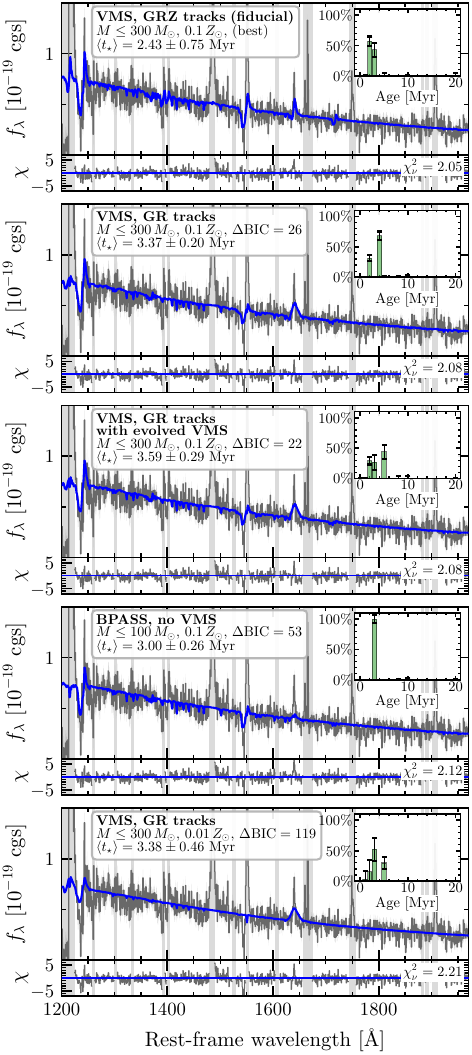}
    \caption{Best-fit stellar population models explored in this work. We find a best-fit stellar population of 0.1\Zsun{} which includes VMSs of up to 300~\Msun{}, with GRZ tracks preferred to GR tracks ($\Delta\rm BIC = 22$; see Sect.~\ref{sec:ficus_analysis}). The models with VMS are preferred over BPASS non-VMS populations ($<100$~\Msun{}; $\rm\Delta BIC = 53$). The bottommost panel shows a fit to a stellar population including VMSs with a lower metallicity of 0.01\Zsun{}, used for the emission line analysis in Sect.~\ref{sec:cloudy_line_modeling}. All the models prefer very young ages of 2.5--3.5~Myr. See Fig.~\ref{fig:pcygni_comparison} for a zoom in relevant features.}
    \label{fig:ficus_models_all}
\end{figure}

\begin{figure}
    \centering
    \includegraphics[width=\linewidth]{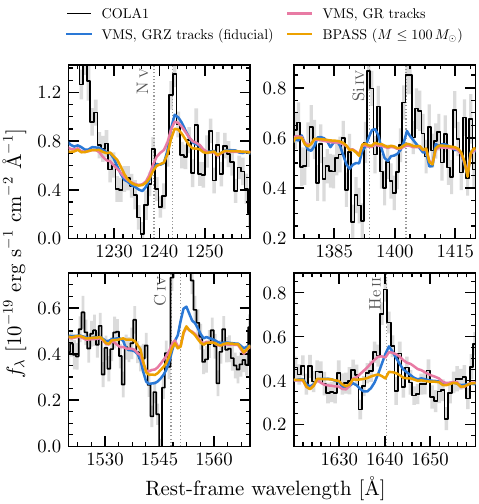}
    \caption{Zoom-ins to the P~Cygni features \ion{N}{v}, \ion{Si}{iv}, \ion{C}{iv}, and the \ion{He}{ii} emission line of COLA1, and three of the models shown in Fig.~\ref{fig:ficus_models_all}. We compare the fits to the stellar populations with VMSs with GR and GRZ tracks (see description in Sect.~\ref{sec:ficus_analysis}), and the BPASS models with stars up to $100$~\Msun{}. The models containing VMS with GRZ tracks provide the overall best-fit across the UV spectrum as shown in Fig.~\ref{fig:ficus_models_all}.
    }
    \label{fig:pcygni_comparison}
\end{figure}

We model the UV spectrum of COLA1 with a combination of single stellar population models (SSP) including VMSs of up to $300$~\Msun{} \citep[][]{Martins2025}.
We use the stellar population fitting code FiCUS \citep[][]{Chisholm2019, SaldanaLopez2025}, following the procedure described by \citet{MarquesChaves2026_vms}. Briefly, FiCUS models the UV spectrum as a linear combination of SSPs, together with the nebular continuum computed with PyNeb \citep[][]{Luridiana2015}. We consider two sets of stellar population models: BPASS models with a maximum stellar mass of $100$~\Msun{}, and the models from \citet{Martins2025}, which include VMSs with masses up to $300$~\Msun{}. All models adopt an extended Salpeter IMF with mass exponent $\alpha=2.35$.

For the VMS models, we explore two prescriptions for stellar mass loss following \citet{Martins2025}: a prescription (VMS\_GR), corresponding to the mass-loss rates inferred for LMC stars, and a metallicity-dependent prescription (VMS\_GRZ), in which the mass-loss rates scale linearly with metallicity relative to the LMC values. We consider combination of SSPs, considering 10 models with ages ranging from 0 to 40~Myr. The nebular continuum is computed with PyNeb assuming an electron density of $n_e=2500$~\unit{cm^{-3}} and an electron temperature of $T=2.2\times10^4$~K. We adopt a \citet{Reddy2015} dust attenuation law. We also use the models in \citet{Martins2025} that include the late stages of evolution of the VMSs. These models explicitly include the advanced stages ($<10$\% of the star lifetimes) where VMSs are predicted to emit increased high-ionization photons, particularly \ion{He}{ii} ionizing continuum that is relevant to assess the nebular emission of \ion{He}{ii} $\lambda1640$ and \ion{N}{v} $1240$, among others \citep[][]{Martins2022, Martins2025}. These models are explored in more detail in Sect.~\ref{sec:cloudy_line_modeling}.

The observed spectrum is fitted in the range 1200--2000~\AA{} assuming a metallicity (of both stars and ISM) of $0.1\,Z_\odot$, matching the inferred gas-phase metallicity of COLA1. Since COLA1 is a known strong leaker of hydrogen ionizing photons \citep[][]{Matthee2018, Torralba2024} we adopt an escape fraction of $\fesclyc = 0.6$, which translates to downscaling the nebular continuum that FiCUS assumes to 60\% assuming a picket-fence escape of ionizing photons. This choice for \fesclyc{} is motivated in Sect.~\ref{sec:ionizing_properties}. Note that a significantly lower \fesclyc{} would produce a strong Balmer jump that is not observed. Our results are robust against small differences in the assumption of the gas metallicity or \fesclyc{}; these would only marginally affect the global $\chi^2$ of all the models due to different infilling of the P~Cygni absorptions from nebular continuum.

We find a best-fit stellar model with mean age $2.43\pm0.75$~Myr, and stellar mass $\approx10^8$~\Msun{} (assuming an IMF with slope $\alpha = 2.35$; \citealt{Salpeter1955}); but we note that $>\,50\%$ of the UV light is emitted by $\sim\,3\times 10^4$ stars with masses $>\,100$~\Msun{}. 
The best-fit models for all the populations tested are shown in Fig.~\ref{fig:ficus_models_all}. The best fit is given by the VMS model with GRZ tracks, followed by the VMS models with GR tracks ($\rm\Delta BIC=22$--26), and performing significantly better than the BPASS models with stars up to only $100$~\Msun{} ($\rm\Delta BIC = 53$). The models including VMS with a low metallicity, $0.01\,\Zsun{}$, perform fairly worse ($\rm\Delta BIC = 119$) due to the lack of absorption features in the stellar spectra. Therefore, the presence of strong P~Cygni profiles suggests that the metallicity of the stars in COLA1 cannot be much lower than 10\% the solar metallicity.

\subsection{Balmer line modeling}\label{sec:balmer_lines_modeling}

\begin{figure}
    \centering
    \includegraphics[width=\linewidth]{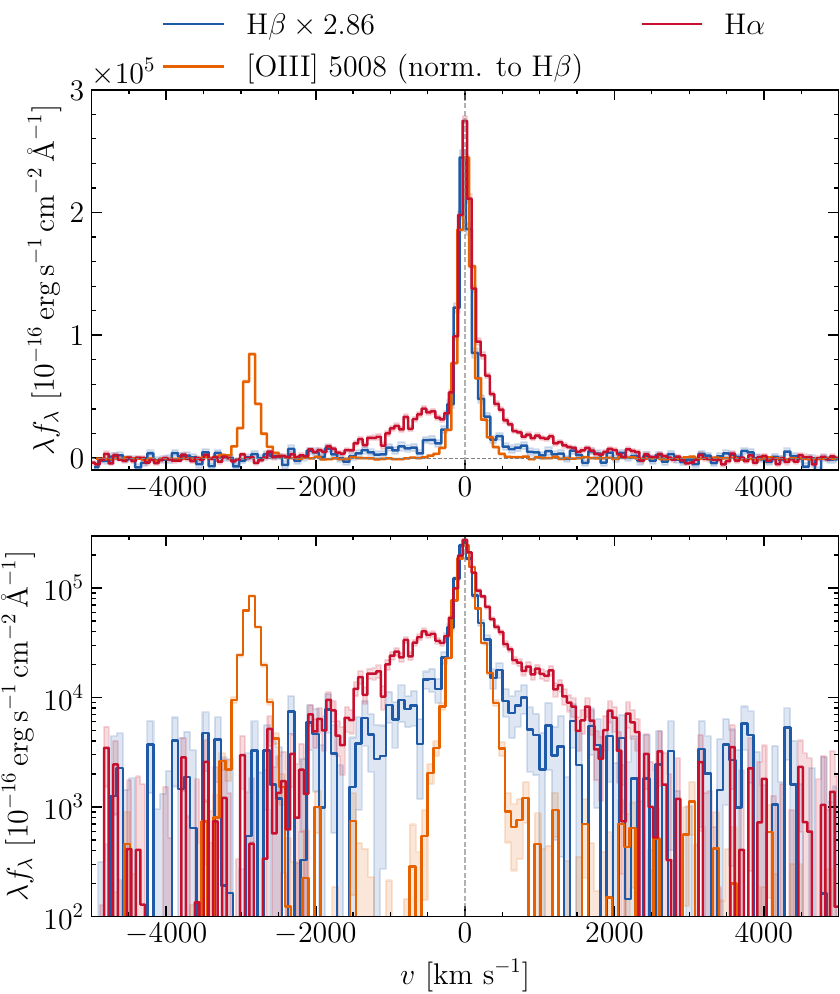}
    \caption{\Halpha{}, \Hbeta{}, and [\ion{O}{iii}] profiles of COLA1 (top: linear scale; bottom: log scale). The \Hbeta{} profile is multiplied by the Case B $\Halpha/\Hbeta$ ratio 2.86, and [\ion{O}{iii}] is normalized to the peak \Hbeta{} for a direct comparison. The narrow emission of the Balmer lines roughly follows Case B with negligible dust attenuation, and show the same width as the forbidden [\ion{O}{iii}] $\lambda 5008$. The broad components of the Balmer lines in turn show a much higher decrement (see also Fig.~\ref{fig:Balmer_decrements}), and \Halpha{} shows a clear blueshifted absorption feature.}
    \label{fig:Ha_Hb_O3}
\end{figure}

\begin{figure}
    \centering
    \includegraphics[width=\linewidth]{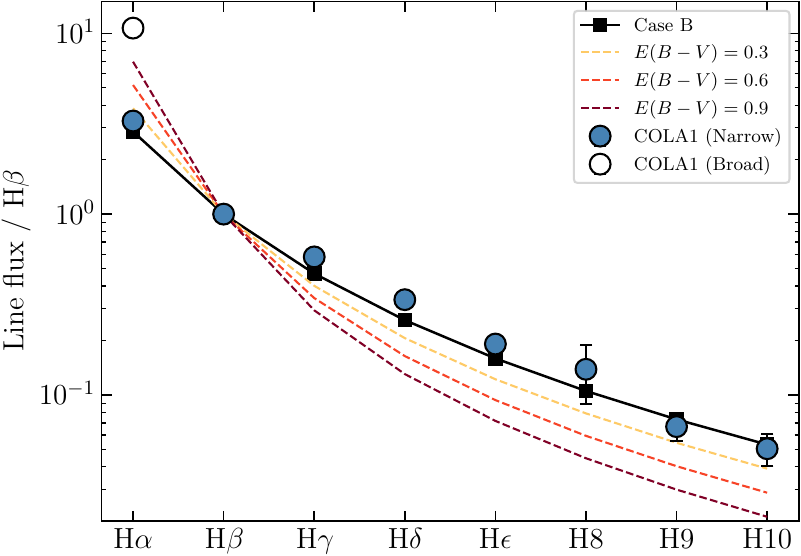}
    \caption{Balmer decrements measured from the narrow-line fit. We show the theoretical Case B values, together with the corresponding attenuated values using the Cardelli extinction law. The measured narrow Balmer fluxes are consistent with Case B recombination with no dust attenuation. The broad components present higher decrements, which could be interpreted as heavy attenuation by nuclear dust, or radiative-transfer effects as is proposed for LRDs (see discussion in Sect.~\ref{sec:balmer_lines_modeling}).}
    \label{fig:Balmer_decrements}
\end{figure}

The \Halpha{} and \Hbeta{} lines of COLA1 show clear broad component emission \citep{Meyer2026_cola1}. In addition, \Halpha{} shows a blueshifted Balmer absorption (see Fig.~\ref{fig:Ha_Hb_O3}), feature that is characteristic of a large fraction of LRDs \citep[e.g.,][]{Matthee2024, Matthee2026, Juodzbalis2026_outflows, Yaganisawa2026, Davis2026}. We fit the \Halpha{} and \Hbeta{} lines with the models described in \citet{Torralba2026_gn9771} and \citet{Matthee2026}, tailored to LRD's Balmer lines. Briefly, our model consists of a narrow component at the line center, an intermediate-width Gaussian, and broad wings described by the intermediate component convolved with a symmetric exponential kernel simulating electron scattering. We also add a Gaussian absorber that affects the broad component and the continuum \citep[see][for a more detailed description]{Matthee2026}. Alternatively, we also fit the same model but replacing the broad exponential wings with a broad Gaussian, and a Lorentzian (Fig.~\ref{fig:Ha_fit_LRD_model}). An exponential broad component is preferred over a Lorentzian ($\rm\Delta BIC = 13$) or a broad Gaussian ($\rm\Delta BIC = 43$), as found in many LRDs \citep[e.g.,][]{Rusakov2026, Scholtz2026}.

Independent of the model choice, we measure the broad and narrow component fluxes of the Balmer lines with the procedure described in Sect.~\ref{sec:emline_measurements}. We fix the intrinsic width of the narrow components of the balmer series from \Halpha{} to H(10-2), and we add an arbitrary number (3) of broad Gaussian components to describe the broad \Halpha{} and \Hbeta{} emission, with the only aim of measuring robust integrated fluxes (reported in Table~\ref{tab:emlines_opt}). Note that for higher-order Balmer transitions no broad component is detected due to sensitivity limitations. In Fig.~\ref{fig:Balmer_decrements} we show the Balmer decrements of the narrow components with respect to \Hbeta{}. The Balmer decrements of the narrow lines across the series are compatible with Case B recombination with negligible dust attenuation (see also Fig.~\ref{fig:Ha_Hb_O3}), compared to the expected decrements using a \citet{Cardelli1989} dust attenuation law.
In turn, the decrement of the broad components is $\Halpha/\Hbeta (\rm broad)\gtrsim 9$, far from the $2.86$ Case B value. The broad components of $\Hgamma$ and successive lines in the Balmer series undetected in our data, possibly due to their low S/N.

\begin{figure}
    \centering
    \includegraphics[width=0.9\linewidth]{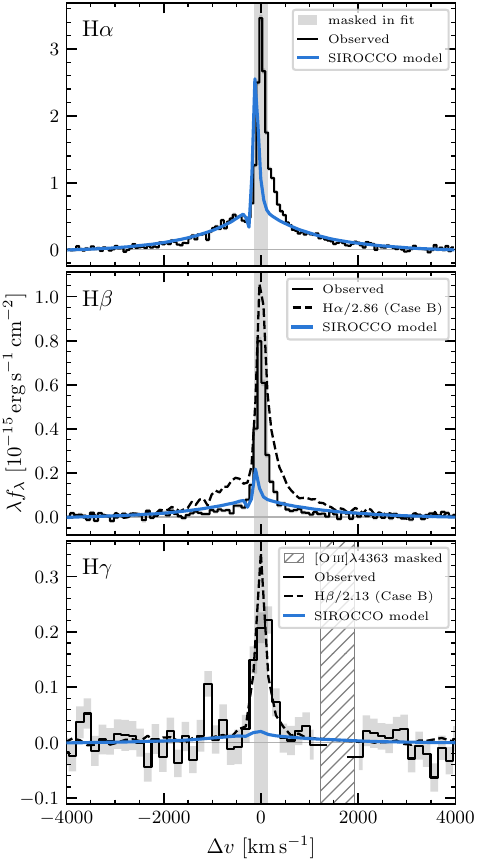}
    \caption{Best-fit Sirocco model to the \Halpha{} broad wings, and the predicted \Hbeta{} and H$\gamma$ profiles. We fit the continuum subtracted \Halpha{} emission line masking the line center, $v_{\rm mask}=-150$ to 500~\unit{km.s^{-1}}, and we show the predicted unscaled \Hbeta{} and H$\gamma$ line from the best-fit model. The Sirocco homologous wind model broadly captures the \Halpha{} tails and P~Cygni-like absorption, and the high Balmer decrement of the broad component ($\Halpha/\Hbeta\sim 8$) without the need for strong dust attenuation. For comparison, in the right panel we show the $1/2.86$ downscaled version of \Halpha{}, which would correspond to a dust-unattenuated Case B recombination line.}
    \label{fig:sirocco_fit}
\end{figure}

The complex broad Balmer line profiles of LRDs could be arising from radiative-transfer effects in a thick wind \citep[e.g.,][but see \citealt{Madau2026} for an alternative interpretation]{Rusakov2026, Naidu2025_BHStar, Naidu2026_etacar, Sneppen2026_sirocco, Torralba2026_gn9771, Chang2026}. We test this hypothesis in order to show that the high Balmer decrement of COLA1 can arise from column density effects in a thick wind rather than localized dust attenuation. For that, we run a grid of Sirocco \citep[][]{sirocco} models \citep[see also][]{Sneppen2026_sirocco, Kaaz2026, Mascia2026}. For simplicity, we simulated a homologous wind, analogous to a spherically expanding layer from a Type IIn supernova \citep[e.g.,][]{Naidu2026_etacar}. with a central blackbody source, and wind metallicity $Z=0.1Z_\odot$. We ran a grid with parameters spanning $\logten(T_{\rm BB,in}/\rm K) \in [4.7, 4.8, 5]$, $\logU \in [0, 0.25, 0.5, 0.75, 1]$, base density $\logten(n_{\rm base} / \rm cm^{-3}) \in [6, 7, 8]$, wind maximum velocity $v_{\max}\in [200, 250, 300, 350]$~\unit{km.s^{-1}}, and column density $\NH\in [24.5, 24.8, 25.0, 25.2, 25.5]$~\unit{cm^{-2}}.
The expanding wind has a density law $n\propto r^{-2}$.
We fit to the broad wings and absorption of the continuum-subtracted \Halpha{} line, after masking the central narrow emission (masked $|v - v_{\Halpha, 0}|<400$~\unit{km.s^{-1}}). For each grid model, the only free parameter for the fit is an arbitrary flux scale $A$.

As shown in Fig.~\ref{fig:sirocco_fit}, the best-fit model is capable of reproducing the broad \Halpha{} wings and blueshifted absorption. That same configuration predicts a \Hbeta{} profile with very weak wings as compared to the Case B recombination expectation, matching the observations. Moreover, this configuration also produces significant narrow emission from the wind with a high decrement, which we interpret as the reason why $\Halpha/\Hbeta$ is higher than expected without the need of any dust attenuation. Indeed, integrating the residuals after subtracting the best-fit model to the observed COLA1 spectrum, the flux ratio is $\Halpha/\Hbeta\sim3$, consistent with the extra emission coming from a different region governed by Case B recombination and no dust attenuation.
While the \Halpha{} absorption is clearly detected, \Hbeta{} does not show it clearly. However, this is expected due to the predicted wind component of \Hbeta{} being weaker, the signal-to-noise is lower, and the absorption could be totally infilled by the strong narrow component.
After finding a best fit model, we ran a finer grid in column density and \logU{} to obtain a best-fit model with $\logten(T_{\rm BB, in}/\rm K)=5$, $\logU=0.2$, $\logten(n_{\rm base}/\rm cm^{-3})=7$, $v_{\max} = 300$~\unit{km.s^{-1}}, and $\logten(\NH /\rm cm^{-2}) = 24.8$ (shown in Fig.~\ref{fig:sirocco_fit}).

\subsection{Nitrogen and Carbon abundance}\label{sec:abundances}

In the following sub-sections we estimate the relative abundances of oxygen, nitrogen, and carbon in the ISM of COLA1. We use two complementary approaches: a single-zone method based on PyNeb line emissivities (Sect.~\ref{sec:abundance_single}), and a two-zone density photoionization model using Cloudy (Sect.~\ref{sec:abundance_two}).

\subsubsection{Single-zone CNO abundances}\label{sec:abundance_single}

The low ratio $\ion{N}{iv}]$ $1483/1486 \approx 0.09$ implies an electron density of $\eden\gtrsim 10^{6.5}$~\unit{cm^{-3}} \citep[e.g.,][]{Keenan1995, Curti2025}, while the high $\ion{C}{iii}] 1906/1909 \approx 1.7$ needs $\eden \lesssim 10^{3.5}$ \citep[e.g.,][]{Schmidt2021}. This suggests that the ISM of COLA1 is composed of a two-zone density system, which could have different chemical abundances abundances, as found for other young starbursts in the literature \citep[e.g.,][]{Martinez2025, ArellanoCordova2026, Chen2026_gnz11}.

We first measure the total abundances using a single-zone density model for direct comparison with other sources in literature. We estimate the nitrogen and carbon abundances from the UV emission line fluxes by comparing to the emissivities from PyNeb \citep[][]{Luridiana2015}, adopting $T_{\rm e}=2.2\times 10^4$~K, and electron density $\eden = 10^4$~\unit{cm^{-3}} as estimated from the optical oxygen lines (see Sect.~\ref{sec:metallicity}); the results depend weakly on these parameters. We use the narrow \ion{N}{iv}] $\lambda 1486$ flux as a tracer of $\rm N^{+3}$, the total flux of the \ion{N}{iii}] $\lambda1750$ quintuplet for $\rm N^{+2}$, \ion{C}{iv} $\lambda\lambda 1548,1551$ for $\rm C^{+3}$, and \ion{C}{iii}] $\lambda\lambda 1906,1909$ for $\rm C^{+2}$. \ion{C}{iv} has a stellar component to the flux, but subtracting the P~Cygni profile of the best-fit stellar population from the observed data only affects the measurements by $\sim 0.01$~dex, well within the 1$\sigma$ errors. From the observed positive line fluxes we obtain $$\logNO = \logten \rm [(N^{+3}+N^{+2}) / O^{+2}] = -0.07^{+0.04}_{-0.05}\,,$$ $$\logCO = \logten \rm [(C^{+3}+C^{+2}) / O^{+2}] = -0.64^{+0.11}_{-0.09}\,,$$ hence $$\logNC = +0.58^{+0.12}_{-0.10} \rm\quad(Total)\,.$$ If we use instead a higher density of $10^{6.5}$~\unit{cm^{-3}} as the \ion{N}{iv}] doublet suggests, the measured abundances change by only roughly 0.01~dex. We have assumed an ICF of $\rm O^{+2}/O = 1$ \citep[see e.g.,][]{Martinez2025}. The nitrogen ionic population of $\rm N^{+3}$ dominates over $\rm N^{+2}$, although we have disregarded $\rm N^{+4}$ and higher states, assuming that lines are powered only by collisional excitation of the hot photoionized gas. We detect some nebular \ion{N}{v} $\lambda1240$ emission (probably photoionized by >77~eV photons), therefore our nitrogen abundance might be conservative (see Sect.~\ref{sec:cloudy_line_modeling}).

\subsubsection{Two-zone CNO abundance photoionization model}\label{sec:abundance_two}

\begin{figure*}
    \centering
    \includegraphics[width=\linewidth]{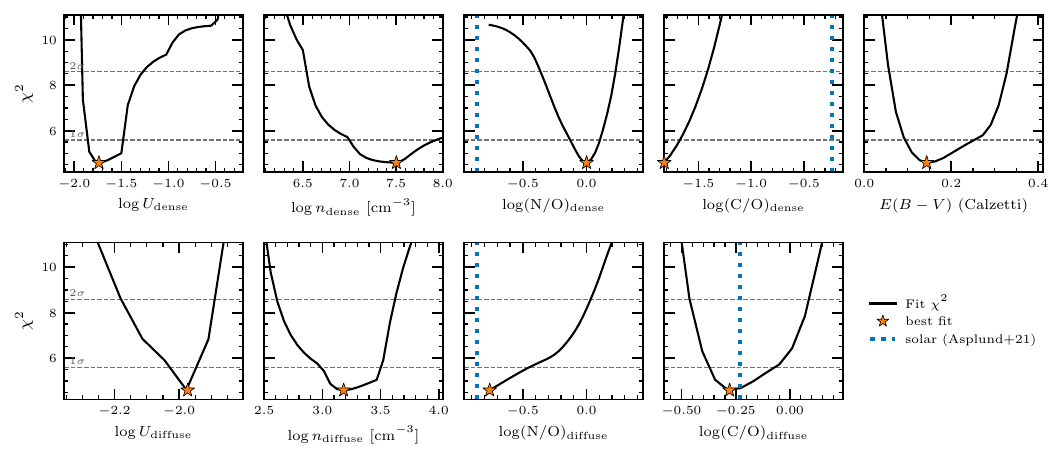}
    \caption{Profiles of $\chi^2$ for the two-zone fit to the UV-to-optical emission lines. These are computed by fixing each parameter to an array of values, and re-fitting for the rest of the parameters to find the best $\chi^2$ (black lines). The best-fit value of each parameter is marked as a yellow star. The C/O abundance of the dense zone, and the N/O of the diffuse zone are railed to the lower values of the grid due to the loose constraints on \ion{C}{iii}] $\lambda 1909$ and [\ion{N}{iv}] $\lambda 1483$, respectively. We mark as horizontal dashed lines the values corresponding to 1$\sigma$ and 2$\sigma$ confidence intervals. We also show the \citet{Asplund2021} solar abundances with a dotted blue line.}
    \label{fig:chi2_profiles_NC}
\end{figure*}

\begin{figure}
    \centering
    \includegraphics[width=\linewidth]{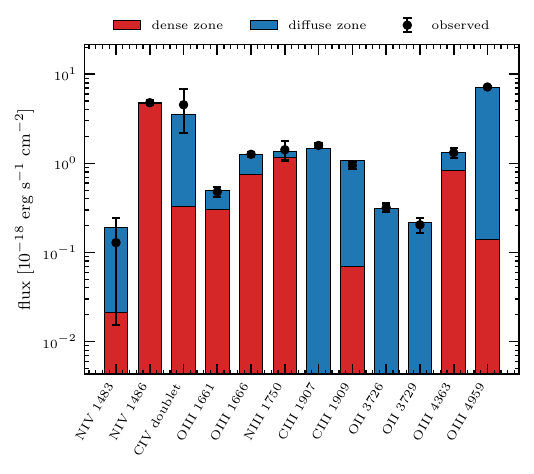}
    \caption{Fluxes of all the CNO emission lines used for the fit of a two-zone abundance model to Cloudy emissivities. We show the decomposition of the total flux produced by the dense (red) and diffuse (blue) zones. The fit is performed for relative fluxes, scaled arbitrarily.}
    \label{fig:line_ratios_fit}
\end{figure}

{Now turning to a two-zone approach, we fit the relative fluxes of all the C, N, and O emission lines across the UV-to-optical spectrum to emissivities using the photoionization code Cloudy \citep[last described in][]{Gunasekera2025}. We ran a Cloudy grid of slabs with metallicity $\rm [M/H]=-1$ and default abundances, density $\logten(\Hden\rm/cm^{-3})$ from 2.5 to 8 in 0.5~dex steps, ionization parameter ranging $\logU \in [-3, -2.5, -2, -1.5, -1, -0.5, 0]$. We also vary the relative N and C abundances in a grid of multiplicative factors $\rm\times N \in [1, 2, 4, 6, 8, 12, 16, 20, 25]$ and $\rm\times C \in [0.03, 0.05, 0.1, 0.2, 0.4, 0.7, 1, 2]$. This is motivated by multiple recent works finding anomalous N/C abundance patterns regularly in high-$z$ galaxies \citep[e.g.,][]{Rai2026, Isobe2025, Shi2026, Berg2026}. We run Cloudy with our best-fit stellar population SED fitted to the UV  as an input (Sect. \ref{sec:ficus_analysis}). We use the stopping condition \texttt{stop efrac -1}, to simulate a radiation-bounded cloud. We fit the relative fluxes of the main CNO lines of COLA1, measured as described in Sect.~\ref{sec:emline_measurements}, using only the narrow components of each line. We use the total positive flux of \ion{C}{iv} without splitting between broad and narrow components due to its uncertain radiative transfer.
We fit the relative line fluxes as a linear combination of any two Cloudy models of the grid, fitting for respective weigths $0<w<1$.
We also use a \citet{Calzetti2000} dust attenuation curve, fitting for $E(B-V)\in[0, 0.4]$.
We find the best-fit using a differential evolution algorithm \citep[][]{Storn1997} implemented in the Python package \texttt{lmfit} \citep[][]{lmfit}. In order to estimate the uncertainty of the best-fit parameters, we perform a $\chi^2$ profiling by fixing each paramenter in an array of values, and re-fitting for the rest of them, finding the best minimum $\chi^2$ for a fixed parameter. We show the $\chi^2$ profiles for each parameter, and the fitted emissivities per density zone for the modeled lines in Figs.~\ref{fig:chi2_profiles_NC} and \ref{fig:line_ratios_fit}, respectively.

The model again converges to two different zones with density $\logten (n / \rm cm^{-3}) = 7.5^{+0.43}_{-0.5}$ and $\logten (n / \rm cm^{-3}) = 3.18^{+0.32}_{-0.19}$.
We find a best-fit dust attenuation parameter $E(B-V) = 0.14^{+0.11}_{-0.05}$, corresponding to a very low dust obscuration. While no dust attenuation is preferred by the narrow Balmer decrements (Fig.~\ref{fig:Balmer_decrements}), this value could be accomodated especially considering a complex dust geometry around the line-emitting regions.
We estimate a nitrogen abundance of $\logNO_{\rm dense} = 0.00_{-0.10}^{+0.13}$ for the dense zone.
This zone requires a very low \logCO{} in order to not overproduce the \ion{C}{iii}] $\lambda 1909$ component, so we can only estimate an upper limit to this quantity $\logCO_{\rm dense} < -1.4$ (2$\sigma$), yielding $\logNC_{\rm dense} > 1.3$ (2$\sigma$; see Fig.~\ref{fig:chi2_profiles_NC}).
The \ion{C}{iv} doublet should contain non-negligible contribution from stellar origin (see Fig.~\ref{fig:ficus_models_all}), and due to its resonant nature, the RT effects are uncertain for this line. However our key results do not strongly depend on a precise measurement of \ion{C}{iv}, since our models predict that most of the \ion{C}{iv} emissivity is produced in the diffuse zone (see Fig.~\ref{fig:line_ratios_fit}).
In turn, for the diffuse zone the nitrogen abundance is unconstrained, given that the \ion{N}{iv}] $\lambda 1483$ component is undetected and blended with the broad component, hence we measure $\logNO_{\rm diff} < -0.45$ (2$\sigma$), while the carbon abundance is well defined by the \ion{C}{iv} and \ion{C}{iii}] lines; thus $\logCO_{\rm diff} = -0.28^{+0.21}_{-0.09}$ and $\logNC_{\rm diff} < 0.02$ (2$\sigma$).} All the measured single- and two-zone abundances are listed in Table~\ref{tab:abundances}.

\begin{table}
\centering
\caption{Measured nitrogen and carbon abundances from the best-fit two-zone model.}
\label{tab:abundances}
\begin{tabular}{lccc}
\hline
Ratio & Dense zone & Diffuse zone & Integrated \\
\hline
$\log_{10}(\mathrm{N/C})$ & $>+1.3$ & $<+0.02$ & $+0.58^{+0.12}_{-0.10}$ \\
$\log_{10}(\mathrm{N/O})$ & $0.00^{+0.13}_{-0.10}$ & $<-0.45$ & $-0.07^{+0.04}_{-0.05}$ \\
$\log_{10}(\mathrm{C/O})$ & $<-1.4$ & $-0.28^{+0.21}_{-0.09}$ & $-0.64^{+0.11}_{-0.09}$ \\
\hline
\end{tabular}
\tablefoot{
Upper or lower limits are given for a 2$\sigma$ confidence level.
Dense and diffuse zone abundances are obtained through the two-zone cloudy emissivity model, and the integrated independently from PyNeb emissivities (see Sect.~\ref{sec:abundances}).
}
\end{table}

\subsection{UV emission line photoionization modeling}\label{sec:cloudy_line_modeling}

In Sect.~\ref{sec:ficus_analysis} we fitted the UV continuum using stellar population synthesis spectra, however we do not include nebular emission lines in our FiCUS models. We fit the UV lines of COLA1 separately. We then use these values as the starting point for an MCMC analysis, in which the model grid is interpolated to evaluate the likelihood at arbitrary parameter values. This allows us to explore the parameter space continuously and obtain a more precise estimate of the best-fitting parameters, together with robust uncertainties.

The spectral fit is performed in the range of rest-frame wavelengths 1200--1950~\AA{}. We mask the \ion{C}{iv} and \ion{Si}{iv} emission lines in this process, as they present P~Cygni stellar wind features that could also be impacted by interstellar absorption, the latter processes not captured by these Cloudy models. As shown in the first panel of Fig.~\ref{fig:cloudy_emlines}, the best-fit using our fiducial stellar population as an input for the Cloudy runs is capable of reproducing the main UV nebular lines (\ion{N}{iv}], \ion{O}{iii}], \ion{N}{iii}], \ion{C}{iii}]). 
The two-zone model converges to a high-density component with a very elevated \logNC{} and a more diffuse zone with less extreme abundances, to reproduce the observed \ion{N}{iv}] and \ion{C}{iii}] ratios.
The broad \ion{He}{ii} $\lambda 1640$ component arises from the stellar SED, but there is a lack of \ion{He}{ii}-ionizing photons from this stellar population model to power the relatively strong narrow \ion{He}{ii} (IP 54.42~eV), which we interpret as nebular. Moreover, the narrow component of the \ion{N}{v} $\lambda\lambda 1240$ doublet (IP 77.47~eV) is also under-estimated, requiring an ionizing spectrum with a large amount of high-energy photons.

We also also consider the possibility that there is an additional source of ionizing photons such as an AGN accretion disk. We run an additional Cloudy grid including an AGN SED\footnote{As defined by the Cloudy AGN command, with the parameters $T_{\rm BB} = 1.5\times 10^5$~K, $\alpha_{\rm ox} = -1.4$, $\alpha_{\rm uv} = -0.5$, and $\alpha_{\rm x} = -1$. See the documentation at \url{https://gitlab.nublado.org/cloudy/cloudy/-/wikis/home}.} mixed with the fiducial stellar model, at fractions of $[0, 12.5, 25, 50, 75]$\% of the UV luminosity at 1500~\AA{}. 
The best-fit model is found with an AGN fraction of 12.5\% (fifth panel of Fig.~\ref{fig:cloudy_emlines}). Including an AGN overshoots \ion{He}{ii} without producing enough \ion{N}{v} flux. These results therefore point towards a very hard ionizing far-UV spectrum, but with a cutoff not exceeding by much the production IP of \ion{N}{v} ($\lesssim 80$ eV).
No other high ionization lines are found across COLA1's spectrum. For example, [\ion{Ne}{v}] $\lambda 3426$ (IP 97.11~eV) is undetected with $\rm EW <1.2$~\AA{} at 3$\sigma$ \citep[compare to e.g.,][]{Chisholm2024, McKaig2024, Reiss2025}. The \ion{Ne}{iv} $\lambda 2424$ is also undetected with $\rm EW < 1.4$~\AA{} at 3$\sigma$, although its IP of 63.45~eV is lower than that of \ion{N}{v}, this line is expected to be luminous in typical AGN-dominated sources \citep[e.g.,][]{Castellano2024, Scholtz2025}. 

\begin{figure}
    \centering
    \includegraphics[width=\linewidth]{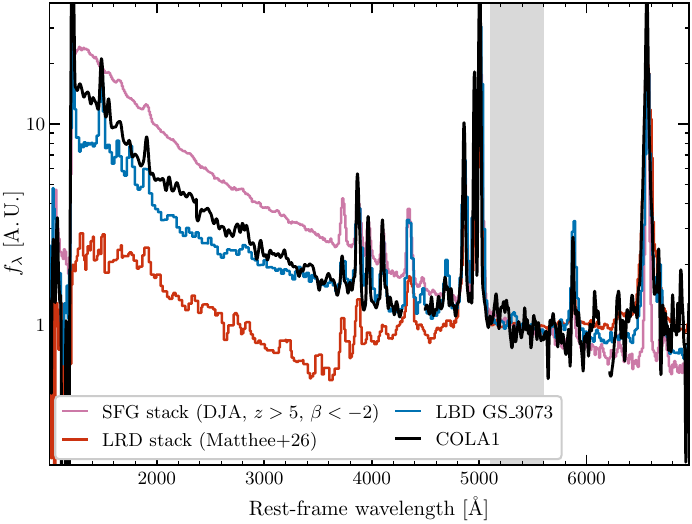}
    \caption{Spectrum of COLA1, from G140M+G235M degraded to PRISM resolution ($R\sim 100$). We compare to the spectrum of a star forming galaxy stack at $z>5$, with $\beta_{\rm UV}<-2$ (pink); an LRD stack (red; \citealt{Matthee2026}); and a prototypical LBD (blue; \citealt{Ubler2023, Brazzini2026}). All the spectra are normalized to the median flux in the optical band marked as a gray shaded region. COLA1 has a very blue UV slope ($\beta_{\rm UV}= -2.5$), but the optical is significantly redder than typical SFGs with similarly blue UV colors.}
    \label{fig:prismified_compare}
\end{figure}

It is well established that stars with lower metallicity produce harder ionizing spectra \citep[e.g.,][]{Schaerer2003}.
We repeat the fitting procedure with models using the 0.01\Zsun{} stellar population model (see Sect.~\ref{sec:ficus_analysis}). In this case, the nebular \ion{He}{ii} is well reproduced, but the fit to this line is noticeably worse, overshooting the broad component. Additionally, this SED is still incapable of producing significant \ion{N}{v} emission, and the overall continuum fit is worse. The models including the advanced phases of VMS evolution of \citet{Martins2025} produce enough high energy photons to power both nebular \ion{He}{ii} and \ion{N}{v}, despite the overall fit being statistically disfavored ($\rm\Delta BIC = 122$). These results highlight the uncertainties of the stellar models of very massive stars, but they show that under certain assumptions these stars could have the ionizing properties needed to power the observed high-ionization UV emission lines \citep{Martins2025}.

\section{Discussion}\label{sec:discussion}

\begin{figure*}
    \centering
    \includegraphics[width=\linewidth]{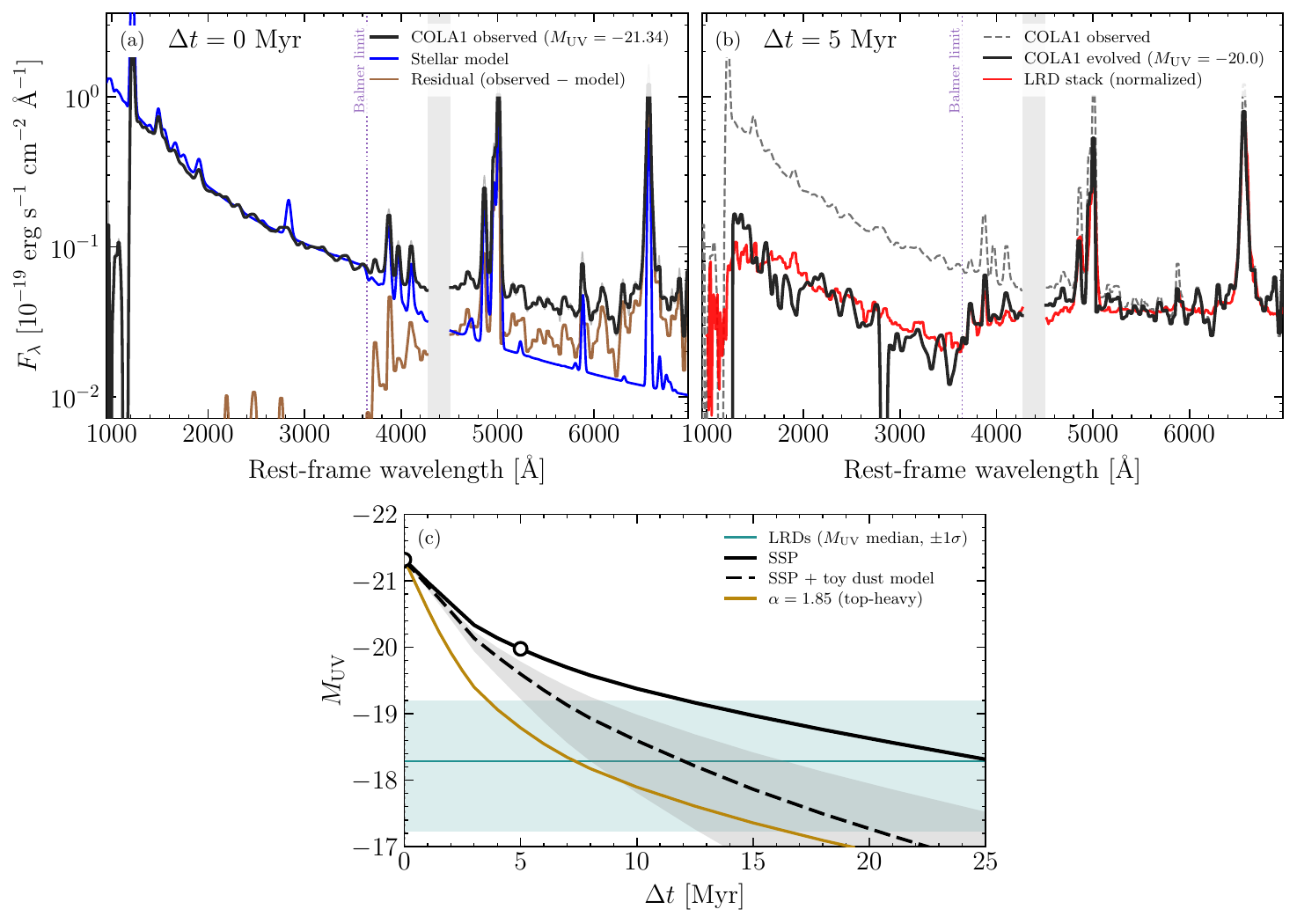}
    \caption{(a) The observed spectrum of COLA1 (black; grating spectra degraded to prism resolution), the best VMS model to fit the rest-UV spectrum (blue), and the subtraction of the two, revealing a residual that remarkably resembles extreme LRDs \citep[e.g.,][]{Naidu2025_BHStar, deGraaff2025}. (b) Aging the best-fit stellar population by 5~Myr leaves a UV-to-optical spectrum (black) that fully resembles that of a typical Little Red Dot, including the characteristic V-shape with an inflection point at the Balmer limit. For comparison, a red solid line shows the spectrum of a stack of LRDs \citep[][]{Matthee2026}. (c) The evolution of the UV luminosity, \MUV{}, of the best-fit stellar population model. A black solid line shows the aging of the fiducial stellar population. We use a toy model to illustrate the effect of SN dust production assuming a screen of 0.02 (0.01--0.04)~\Msun{}/SN at 200~pc (black dashed line and gray shaded area). We also show the evolution assuming a top-heavy IMF with high-mass slope $\alpha = 1.85$ instead of the original high-mass slope $\alpha=2.35$ assumed by the SSP model. We compare to the median and 16,84 percentiles of the spectroscopic sample in \citeauthor{degraaff2025_blackbodies}~(\citeyear{degraaff2025_blackbodies}; teal solid line and shade).
}
    \label{fig:vshape_evol}
\end{figure*}

\subsection{A Little Red Dot outshone by a population of VMSs}\label{sec:LRD_plus_vms}

The UV of COLA1 shows all the classical signatures of a young population of massive stars, while the optical continuum and the luminous broad Balmer lines are typically associated with an AGN or an LRD. The strong \ion{N}{iv}] emission indicates the presence of highly nitrogen-enriched gas \citep[e.g.,][]{Zhu2025, Chen2026_gnz11, Nakane2026, Naidu2026_momz14}. The absence of low-ionization absorption lines (e.g., \ion{Si}{ii}, \ion{O}{i}, \ion{C}{ii}, \ion{Al}{ii}), the strong double-peaked \lya{} emission and weak Balmer jump are indications of a high ionizing continuum escape fraction \citep[e.g.,][]{Izotov2018, Keerthi2026, Nakane2026}.

The optical continuum of COLA1 cannot be described with the same SSP model that reproduces the UV, requiring an extra component.
In Fig.~\ref{fig:prismified_compare} we show the complete G140M+G235M spectra, degraded to the resolution of NIRSpec/PRISM ($R\sim 100$). We compare to the NIRSpec/PRISM stack of LRDs from \citet{Matthee2026}, and to the prototypical ``Little Blue Dot'' GS\_3073 \citep[][]{Ubler2023, Brazzini2026}. COLA1 has a redder optical slope than the LRD stack, and the overall UV-to-optical continuum is more similar to that of GS\_3073. However, a critical difference is that COLA1 shows Balmer absorption in the \Halpha{} line, which LBDs lack. Other notable differences can be seen between the general UV shapes of COLA1 and GS\_3073, with the latter showing relatively stronger UV emission lines \citep[see][]{Ji2024}. We also compare to a median stack of star forming galaxies (SFGs), selected from the JWST Dawn archive by imposing a secure visual inspection redshift $z>5$, and blue UV slopes $\beta < -2$. As one can see in Fig.~\ref{fig:prismified_compare}, the optical continuum of typical blue SFGs remains blue, while COLA1 shows a slight turnover, that can be interpreted as an extra component in addition to the young stellar light.

The residual of subtracting the best-fit stellar model (fit only to the UV) from COLA1's spectrum reveals a continuum component that dominates the rest-optical (first panel of Fig.~\ref{fig:vshape_evol}), and has a turnover near the Balmer limit ($\rm H_{n=2\rightarrow\infty}; 3645$~\AA{}). This component could in principle be identified as an older stellar population with a Balmer break, or a dust-obscured AGN. However, we find a more compelling interpretation that the residuals trace a red, compact component, such as the optical continuum of an LRD. We favor this interpretation because we also detect broad Balmer emission lines with extreme Balmer decrements, as well as blueshifted H$\alpha$ absorption---the latter being one of the defining properties of LRDs rarely seen in any population of AGN \citep[e.g.,][]{Matthee2026, Juodzbalis2026_outflows, Yaganisawa2026, Shangguan2026}.
Some works have shown that the characteristic UV-to-optical ``V-shape'' of many LRDs can be decomposed into a blue star-forming galaxy component and a red source similar to the red component we find here dominating the optical light \citep[e.g.,][]{Naidu2025_BHStar, Ji2025_blackthunder, Sun2026, degraaff2025_blackbodies, Weibel2026}. We propose that the spectrum of COLA1 can be well described by the same scenario, except that here we catch the stellar population at an exceptionally young age.

\subsection{Evolution of the stellar population -- COLA1 as a rare snapshot of an emerging LRD}\label{sec:vms_evolution}

Due to the steep stellar mass-luminosity relation \citep[e.g.,][]{Russell1923, Hertzsprung1923, Eker2015}, the UV spectrum of COLA1 is dominated by the light of the most massive stars, which have a short lifetime \citep{Martins2025}. If we assume no significant star formation occurs after the observed burst, the UV will fade as the most massive stars end their lives in a timescale of 2--5~Myr. We test this directly by evolving our best-fit stellar population. In the top right panel of Fig.~\ref{fig:vshape_evol} we show the stellar model evolved by $\Delta t=5$~Myr into the future.
In order to build the evolved spectrum, we used the SSP models that correspond to the ages of the two starburst of our best fit (see Fig.~\ref{fig:fig1}) plus 5~Myr, with the same initial mass, weighted by the observed light fractions at the present epoch, essentially preserving the star-formation history. That is, we replace the population at 3~Myr by one at 8~Myr, and since 7~Myr is not available in our library, we linearly interpolate the models with 5~Myr and 8~Myr. We also add nebular continuum and emission lines by giving the evolved SSPs as a input to Cloudy, with the same parameters as the fiducial $\Delta t=0$ model, and re-scaling the ionization parameter by the decrease of ionizing continuum ($Q_{\rm H^+}$) in the integrated stellar population model SEDs. The bottom panel of Fig.~\ref{fig:vshape_evol} shows the \MUV{} evolution of our best-fit SSP model up to 25~Myr.

The UV luminosity of the aged stellar population fades as the most massive stars disappear. After $\sim 5$~Myr, our evolutionary toy model will leave a UV-to-optical spectrum that fully resembles an LRD with a ``V-shape'' ($\beta_{\rm UV, aged} = -1.7\pm 0.1$ and $\beta_{\rm opt, aged} = +0.1\pm0.1$), and would pass standard LRD selection criteria \citep[e.g.,][]{degraaff2025_blackbodies, Hviding2025, Kocevski2026}. Here we assume that the lifetime of the red component is longer than that of the VMSs and that it does not evolve significantly during this process, although in a realistic scenario some evolution might be expected (see Sect.~\ref{sec:interpretation} for a more detailed discussion about timescales).

The appearence of dust can make the \MUV{} decline over time even steeper. We illustrate this with a very simple toy model of dust production by SNe. We assume that every $>8$~\Msun{} star undergoes a supernova at the end of its life producing 0.02~\Msun{} of net dust mass, after reverse-shock destruction \citep[][]{Bianchi2007}. In each time step we add this dust mass to a spherical shell with radius 200~pc (roughly the UV size of COLA1). The UV attenuation is estimated using a mass absorption coefficient of $\kappa_{1500} = 5\times10^4$~\unit{cm^{2}.g^{-1}} \citep[for a Milky Way attennuation law, $R_v = 3.1$;][]{Weingartner2001}. We show the effect of applying this dust model evolution as a dashed line in the bottom panel of Fig.~\ref{fig:vshape_evol}.

The UV fading also depends heavily on the IMF. The SSP models that we employ assume a slope of $\alpha = 2.35$ \citep[][]{Salpeter1955, Martins2022, Martins2025}.
To illustrate the effect of a top-heavy IMF we adopt a parametrised description to modify the available SSP. In this toy model, we assume that the starburst fades as the time-dependent main-sequence turnoff mass $M_{to}(t)$ decreases. Therefore the UV luminosity can be approximated as
\begin{equation}
L_{\rm UV}(t,\alpha) \propto \int_{0.1M_\odot}^{M_{\rm to}(t)} m^{k-\alpha}{\rm d}m\,,
\label{eq:LUV_law}
\end{equation}
with $\alpha$ the IMF slope and $k$ the effective luminosity–mass exponent, $L_\star \propto M^{k}$. We use as main-sequence turnoff masses the H-burning end ages from \citet{Ekstrom2012}.

The exponent $k$ is calibrated requiring the expression above to reproduce the measured fading of the BPASS models for the utilized  slope $\alpha = 2.35$ yielding $k = 2.5$. Then we compute the evolution of \MUV{} from Eq.~\ref{eq:LUV_law}, normalizing to the observed value at $t=0$. The evolution of \MUV{} of this toy model is shown in the bottom panel of Fig.~\ref{fig:vshape_evol} as a gold solid line, assuming a top-heavy IMF with $\alpha = 1.85$ \citep[e.g.,][]{vandokkum2024}. For this assumption, the starburst reaches $\MUV = -20$ in only $\sim 2$~Myr, much faster than the 5~Myr of the fiducial SSP assuming a Salpeter-like IMF.
Note that hydrodynamical simulations of very dense star clusters predict that, due to converging mass flows, the most massive stars can grow efficiently, and extreme IMFs can be achieved up to log-flat shapes \citep[$\alpha=1$;][]{Chon2026_sms_cluster}. Generally, the fading process to a typical LRD UV magnitude of $\sim-19$ can happen faster than the duty cycle of $\sim 10$--20~Myr, inferred from the incidence of LRDs amongst $\sim 10^8$~\Msun{} galaxies \citep[][]{Sun2026}, suggesting that the red engine may outlive the VMS population.

\subsection{The origin of the high N/C abundance}\label{sec:NC_yields_model}

In Sect.~\ref{sec:abundances} we have reported a very high single-zone N/O abundance, only comparable to the most enriched systems at high-$z$ \citep[e.g.,][]{Ji2026_NO, Naidu2026_momz14, Morel2026}. Nitrogen is a bottleneck element in the stellar CNO cycle. Convection and rotation of massive stars can efficiently cause mixing in the stellar envelope, bringing the N-rich, C-poor material of the H-burning core to the surface, allowing it to be ejected through stellar winds \citep[e.g.,][]{Maeder2000}. For this reason, the N/C abundance ratio has the potential to be a strong indicator of enrichment by stellar yields. Below we explore models of yields by rotating VMSs' winds, massive Pop\,III stars (up to $\sim 6000\ \Msun$), and SMS nucleosynthesis ($\gtrsim 10^4\ \Msun$).

\begin{figure}
    \centering
    \includegraphics[width=\linewidth]{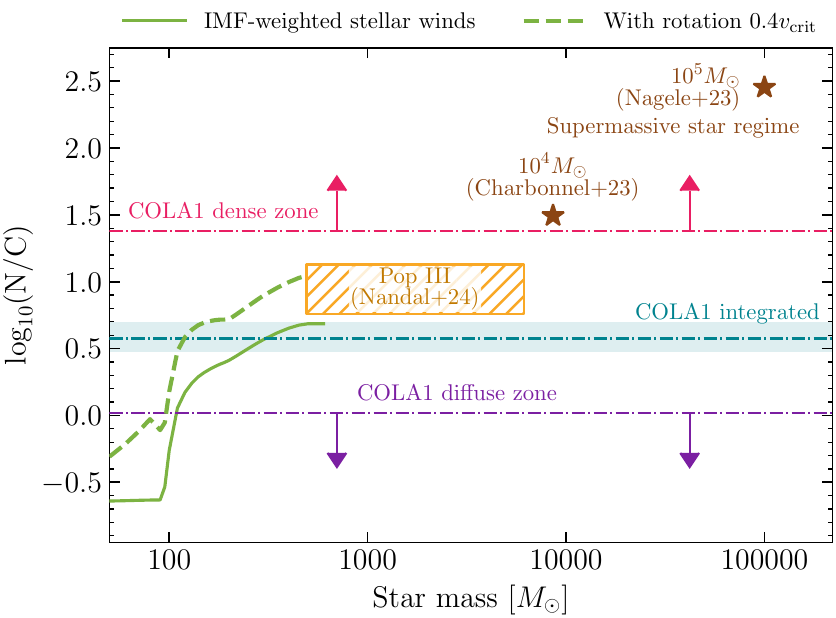}
    \caption{N/C abundance of the different zones modeled to explain COLA1's spectrum, compared to different yield models of stellar populations with different masses. We show the abundances from stellar winds assuming no ISM mixing for a $0.14$\Zsun{} stellar population with a \cite{Kroupa2001} IMF and increasing high-mass cutoff (green solid line), whereas stellar models including fast rotation are shown as a green dashed line ($0.4\,v/v_{\rm crit}$; \citealt{Higgins2025}). As a yellow hatched region, we show the yields of Pop\,III stars of up to $\approx 6000$~\Msun{} of~\cite{Nandal2024}. Finally, we show the abundances of SMS nucleosynthesis as upper limits for 0.1\Zsun{} models of $10^4$~\Msun{} \citep[][]{Charbonnel2023} and $10^5$~\Msun{} \citep[][]{Nagele2023}. Only the latter two models can match COLA1's most dense gas N/C abundances.}
    \label{fig:logNC_yields}
\end{figure}

\subsubsection{VMS winds}

We computed the \logNC{} abundances of the stellar wind yields of the PARSEC VMS models, which include stellar tracks for masses up to 600~\Msun{} \citep[][]{Costa2025, Nguyen2025}.
In Fig.~\ref{fig:logNC_yields} we show the abundance as a function of an increasing IMF high-mass cutoff. The total yields of the stellar population depends on age, so we plot the yields for our fiducial best-fit population of 2--3~Myr, weighted by total mass, but note that these ages are close to what would yield the maximum N/C achievable by a VMS population \citep[see e.g.,][]{Shi2026}.
We also consider the ejecta of fast-rotating 0.14\Zsun{} VMSs \citep[$v/v_{\rm crit} = 0.4$;][]{Higgins2025}.

In Fig.~\ref{fig:logNC_yields} we compare the \logNC{} abundances of COLA1 (dense zone, diffuse zone, and integrated) with the yields of VMS populations.
Achieving the measured single-zone $\logNC = +0.58$ with such stellar populations would require an IMF cutoff of at least $\approx 300\ \Msun$ for non-rotating stars, and $\approx 100\ \Msun$ for the case of $v/v_{\rm crit}=0.4$ rotating stars.
However, note that the green curves plotted in Fig.~\ref{fig:logNC_yields} assume zero dillution with the ISM metals; only net stellar yields are considered. This would correspond to an extreme case where the original intracluster gas has been blown away by a previous generation of stars \citep[see e.g.,][]{Marques-Chaves2024}. Therefore, we conclude that it is unlikely that the measured single-zone N/C abundance can be produced by the winds of VMS only.

Turning to the two-zone aboundance estimates, the dense-zone $\logNC{}>1.3$ simply cannot be reached with this models with an IMF cutoff of up to $\gtrsim 600$~\Msun{}, even with fast-rotating stars.
In order to explain the observed \logNC{} in our toy model we would need $\gtrsim 50\%$ of the stars rotating at an extreme $v/v_{\rm crit}=0.7$ \citep[][]{Meynet2010}, which renders unfeasible compared to the observed rotation of indiviudal massive stars in the local Universe \citep[e.g.,][]{Harries1998, Ramirez-Agudelo2013}.

\subsubsection{Pop\,III stars}

Alternative mechanisms need to be invoked to explain the extreme \logNC{} values in COLA1. \citet{Nandal2024} predicted that the ejecta of Pop\,III stars of up to $\sim 6000$~\Msun{} would be composed mainly of H, He and N. In Fig.~\ref{fig:logNC_yields} we show the \logNC{} prediction by \citet{Nandal2024} as a result of the Pop\,III ejecta in the embryonic halos of the \citet{Regan2020} hydrodynamical simulation. The abundances of the Pop\,III ejecta of this model broadly match the integrated value of COLA1. The upper and lower limits of the yellow box correspond to the assumptions that the 10\% and 100\% of the material above the CO core in the Pop\,III stars is ejected, respectively. Note that the integrated ejecta depends strongly on the assumption of how the IMF is sampled; higher mass stars produce higher N/C in general. As discussed in Sect.~\ref{sec:ficus_analysis}, the best-fit stellar population to the UV continuum needs a metallicity not significantly lower than $0.1\,\Zsun{}$ to produce the observed metal absorptions. However, it could be possible that the observed N/C abundances were originated from Pop\,III nucleosynthesis, and the currently observed stellar population is a second generation of stars. In this case, the metals traced by the stellar absorption features and the nebular lines (e.g., [\ion{O}{iii}]) must have been produced by an earlier population, most likely through core-collapse supernovae, whose C- and O-rich ejecta would dilute the N/C ratio of the Pop\,III material. The young age of the observed population ($\sim2.5$~Myr) implies that it has not yet had time to alter the gas through its own AGB stars \citep[$\gtrsim30$~Myr;][]{Karakas2010} or core-collapse supernovae of the current population \citep{Pipino2009}. In addition, the nitrogen excess appears to be concentrated in the dense zone of our two-zone model (Sect.~\ref{sec:abundances}; see also Fig.~\ref{fig:chi2_profiles_NC}), which favours a recent, localized enrichment over a pre-existing one that would have mixed throughout the ISM.

\subsubsection{Supermassive stars}

Supermassive stars (SMS; $\sim 10^4$--$10^5$~\Msun{}) have been proposed as candidates responsible for the elevated N/O enrichment through hot CNO nucleosynthesis in dense stellar clusters and high-$z$ galaxies \citep[][]{Nagele2023, Charbonnel2023, Marques-Chaves2024}, or the anomalous chemical enrichment in globular clusters \citep[][]{Carretta2009a, Carretta2009b, Denissenkov2014}.
\citet{Charbonnel2023} modeled the nucleosynthesis of a fast growing SMS reaching $10^4$~\Msun{} using the stellar evolution code MESA and following the prescriptions of \citet{Fragos2023}. In Fig.~\ref{fig:logNC_yields} we show the N/C abundance of the $\rm [Fe/H] = -1.14$ model of \citet{Charbonnel2023} without mixing with the intracluster gas as an upper limit. The very high maximum $\logNC\approx 1.5$ of this model is compatible with the dense-zone abundance of COLA1, albeit with the limiting case of assuming no mixing with the cluster gas or pollution from other metal-enrichment sources.
\citet{Nagele2023} presented a model using the code HOSHI and a \citet{Yoshida2018} mass loss prescription. We also show in Fig.~\ref{fig:logNC_yields} their $10^5$~\Msun{} (close to the theoretical limit due to general-relativistic instabilities; \citealt{Chandrasekhar1964}) as an upper limit assuming no mixing.
As shown in Fig.~\ref{fig:logNC_yields}, SMSs of $10^4$--$10^5$~\Msun{} could easily explain the N/C enrichment after including some amount of dilution with the original intracluster gas. Additionally, the yields of SMSs could explain the extreme N/C abundance of COLA1 in its dense zone.

\subsubsection{Tidal disruption events}

When a star passes too close to a massive black hole, its structure can be torn apart by the gravitational field causing a TDE \citep[e.g.,][]{Lacy1982, Evans1989}. The TDEs yield the contents of the stellar interiors as debris, polluting the ISM with the unprocessed CNO-cycle materials, and could also explain the elevated N/C abundance measured \citep[e.g.,][]{Kochanek2016, Cameron2023}.
Extreme N/C values have been reported for TDEs in the local universe as high as $\logNC>0.9$ \citep[][]{Yang2017}.
In a massive, very dense star cluster containing massive stars, where a IMBH or SMBH is probably also present (see discussion in Sect.~\ref{sec:interpretation} below), an elevated number of TDEs is an expected consequence \citep[e.g.,][]{Sakurai2019, Isobe2025}.
The intracluster gas could be originaly polluted by the winds of the VMS detected in the UV spectrum of COLA1 in first instance, thence TDEs could possibly further boost the N/C abundance. The material disrupted by the TDEs could also be related with to the formation of a dense gaseous envelope around a massive black hole leading to the LRD phenomenon \citep[e.g.,][]{Loeb1997, Metzger2022, Bellovary2025}, although this interpretation may be difficult to reconcile with the weak observed variability on a population level among LRDs \citep[e.g.,][]{Kokubo2025, Liu2026_twinkle, Stone2026}.

\subsection{Broad \ion{N}{iv}] emission and dense outflows}\label{sec:outflows}

COLA1 presents a broad component in multiple emission lines (see Appendix~\ref{app:emline_fluxes}). In particular, we detect broad Gaussian components in the \ion{N}{iv}] $\lambda\lambda 1483,1486$ doublet ($\rm FWHM = 1800 \pm 200$~\unit{km.s^{-1}}, $\Delta v = +200 \pm 200$~\unit{km.s^{-1}}) and \ion{C}{iv} $\lambda1550$ ($\rm FWHM = 1040\pm 50$~\unit{km.s^{-1}}, $\Delta v = -86 \pm 200$~\unit{km.^{-1}})\footnote{Also \ion{He}{ii} $\lambda 1640$, but this one in particular is discussed in detail in Sect.~\ref{sec:cloudy_line_modeling}.} in the rest-UV. The \ion{C}{iv} component is consistent with the systemic redshift, although very uncertain due to the degeneracy with the P~Cygni absorber, and could also have a component due to resonant scattering. The \ion{N}{iv}] doublet broad components in turn are redshifted (moving away from the line of sight), although compatible with being at the systemic velocity within 1$\sigma$ (see Fig.~\ref{fig:all_uv_lines}). Similar outflows were detected in other systems with hints of VMS and similar spectra to COLA1, such as GNz11 \citep[][]{Chen2026_gnz11, Nakane2026_gnz11}. A key difference of our results with respect to GNz11 is that we report broad emission associated with both components of the \ion{N}{iv}] $\lambda\lambda 1483, 1486$ doublet, with the $\lambda 1483$ component being stronger. This is made evident in Fig.~\ref{fig:all_uv_lines}, where one can see a broad tail extending to the blue side of the doublet, not captured by a narrow component. This measurement could in fact be affected by model degeneracies, but at face value it suggests that the \ion{N}{iv}] outflow has a density $\eden \lesssim 10^{6.5}$~\unit{cm^{-3}}. In the rest-optical, we also detect outflow components in the [\ion{O}{iii}] $\lambda\lambda 4960,5008$ doublet ($\rm FWHM = 431 \pm 7$~\unit{km.s^{-1}}, $\Delta v = 53 \pm 3$~\unit{km.s^{-1}}) and [\ion{Ne}{iii}] $\lambda 3860$ ($\rm FWHM = 590 \pm 70$, $\Delta v = 10 \pm 20$). The [\ion{Ne}{iii}] outflow is again consistent with systemic velocity, but the broad [\ion{O}{iii}] $\lambda 5008$ emission is clearly redshifted. The optical outflows are much narrower than the strong and broad \ion{N}{iv}, but the fact that we observe them in several lines at the same time suggests that there is outflowing gas, with density $< 10^6$~\unit{cm^{-3}}. The difference of widths across the line components could reflect a structure in the ionization, and chemical enrichment pattern of the outflow (i.e., N/O and N/C; see Sect.~\ref{sec:abundances}).

\subsection{Physical interpretation -- Formation of a massive seed through runaway stellar mergers}\label{sec:interpretation}

The three defining properties of COLA1---a rest-UV continuum dominated by very massive stars (VMSs), an LRD-like core that dominates the rest-optical emission, and an extreme nitrogen enhancement---may have a causal relation.
It has been argued that many LRDs are well described as the combination of a young starburst and a red compact engine which could be emerging from the same physical process \citep[e.g.,][]{Inayoshi2026_lrd_starburst, Chisholm2026}. 
As an alternative explanation, the red continuum and broad Balmer lines could be tracing a rapid (super-Eddington) accretion phase in a pre-existing massive black hole with significant dust that obscures our line of sight \citep[e.g.][]{Madau2026_se_lrds}, in which case our data still suggest such phenomena likely take place amidst dense star clusters.

Runaway collisions in very dense star clusters can rapidly form IMBHs, with possibly the appearence of an SMS as intermediate step \citep[e.g.,][]{Portegies1999, Gieles2018}. In this scenario, the most massive stars sink to the cluster center and merge repeatedly.
Hydrodynamical simulations show that an SMS can grow to $\sim 10^4$~\Msun{} in timescales of $<1$~Myr through runaway collisions, and can last up to $\sim 5$~Myr through stellar-collision rejuvenation \citep[e.g.,][]{Gieles2018, Vergara2026, Rantala2026, Williams2026}. In most models, the SMS lives as long as it is continously accreting material, until it eventually collapses to an IMBH of $M_{\rm BH}\approx10^3$--$10^4$~\Msun{} due to relativistic instability \citep[e.g.,][]{Woods2017, Hammerle2018}.
In this process, a population of VMSs is expected as a natural by-product. The timescales of SMS formation are comparable to the lifetimes of the VMSs, so both could be present within similar timescales.

Our measured (barely resolved) UV effective radius of $\sim 200$~pc, and compactness with a stellar surface mass density of $\Sigma_{*}\sim 400\ \Msun{}\rm\,pc^{-2}$, match individual star cluster complexes \citep[e.g.,][]{Bouwens2021}. The stellar density is also in line with hydrodynamical simulations where a handful of very dense stellar clusters can form, containing VMSs and SMSs \citep[][]{Rantala2026, Chon2026_sms_cluster, Williams2026}. Our data cannot resolve individual clusters, but the compactness of COLA1 leaves room for such a configuration. Consider for example, the highly lensed system ($\mu\approx 20$) reported by \citet{Yanagisawa2026_lensed}, where LRDs appear to form in compact off-nuclear clusters.
This formation setup offers an explanation for the nitrogen enhancement through VMS stellar winds, TDEs produced by the newly formed IMBH \citep[][]{Sakurai2019}, or the ejecta of an SMS as discussed in detail in Sect.~\ref{sec:NC_yields_model} above.

Now we consider the possibility that the optical emission of COLA1 could be related to an SMS formed within the environment of a dense stellar cluster. SMSs are expected to have inflated, cool envelopes presenting effective temperatures as low as $\sim 4000$~K, and broad permitted lines with P~Cygni profiles in the optical spectrum \citep[e.g.,][]{Gieles2018, Nandal2026, Zwick2026}. Nevertheless, some of the current models of SMS atmospheres struggle to reproduce the red colors of LRDs' optical spectra without invoking heavy dust attenuation \citep[e.g.,][]{Martins2020, Martins2026}. In addition, the short lives inferred for SMSs also challenge an interpretation of the red continuum of LRDs as active SMSs directly. The red optical continuum and broad emission lines that we observe in COLA1 (and in LRDs in general) could more probably represent a stage following an SMS, where the core has collapsed due to gravitational instability as an IMBH seed \citep[e.g.,][]{Inayoshi2026_lrd_starburst, Sun2026_overmassive, Curtis2026, Liu2026_models, Nandal2026_pulsations}.
Before the collapse, the SMS may rapidly lose mass through wind pulsations, providing a possible explanation for the assembly of a very dense cocoon of gas giving rise to the observed LRD features \citep[e.g.,][]{Naidu2026_etacar}. This cocoon formed from the SMS ejecta would be expected to contain N-rich material from the SMS nucleosynthesis \citep[e.g.,][]{Nandal2026_pulsations}, which may explain the high N/O and N/C recently measured in the dense gas ($\eden>10^{6.5}$~\unit{cm^{-3}}) of some LRDs \citep[e.g.,][]{Isobe2025, Ji2026_NO, Tang2026_spurs_NC}.
In this case, accretion into the newly formed massive black hole surrounded by the dense gas is likely to be the powering source of the LRD's bolometric luminosity. The engine may be sustaining the envelope through radiation pressure, emitting at roughly the Eddington ratio \citep[e.g.,][]{Begelman2008, Begelman2026}, and the lifetime of this phase could last for up to $\sim 10$--$100$~Myr, mostly depending on the mass of the system and the uncertain accretion physics \citep[e.g.,][]{Roman-Garza2026, Santarelli2026, Wang2026_LBD_evolution}.

\subsection{Comparison to other LRDs and N-emitters}\label{sec:compare_spurs}

\begin{figure*}
    \centering
    \includegraphics[width=\linewidth]{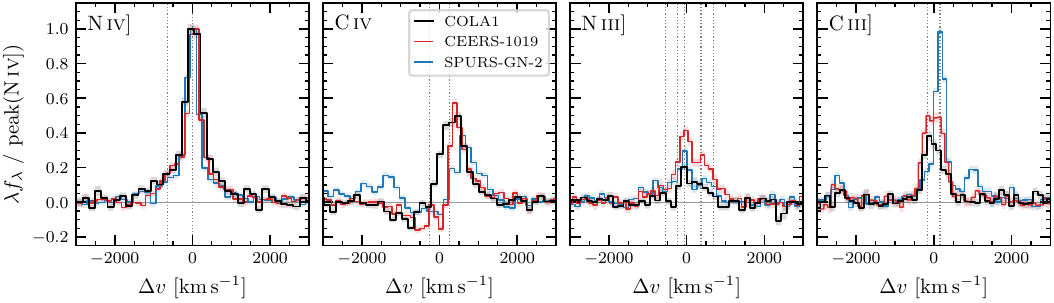}
    \caption{Comparison of \ion{N}{iv}], \ion{C}{iv}, \ion{N}{iii}], and \ion{C}{iii}] emission lines of COLA1 (black); CEERS-1019, a VMS galaxy at $z=8.678$ (red; \citealt{Larson2023, Marques-Chaves2024, MarquesChaves2026_vms}); and the LRD SPURS-GN-2 \citep[][]{Tang2026_spurs_NC}. The spectra have been normalized to the peak of \ion{N}{iv} $\lambda 1486$ and continuum-subtracted. All three show similar \ion{N}{iv}] profiles with broad components. The profile of the resonant \ion{C}{iv} line of COLA1 shows significantly higher flux near systemic velocities, probably related to a high \fesclyc{} \citep[][]{Naidu2022}. The \ion{C}{iii}] doublet of COLA1 is remarkably different, suggesting that this line is mostly emitted in a diffuse gas ($\eden \lesssim 10^{3.5}$~\unit{cm^{-3}}) unlike the other two sources.}
    \label{fig:lines_spurs_comp}
\end{figure*}

In Fig.~\ref{fig:lines_spurs_comp} we compare the main nitrogen and carbon UV lines of COLA1 to the galaxy CEERS-1019 ($z=8.678$; \citealt{Larson2023}) and the LRD SPURS-GN-2 \citep[$z=7.186$;][]{Tang2026_spurs_NC}. These two sources have highly similar \ion{N}{iv}] profiles, with a prominent $\lambda 1486$ component and a broad base, indicative of outflows as discussed in Sect.~\ref{sec:outflows}. They also have high measured N/C abundances: $\logNC = 0.36$, and $\logNC \approx 0.6$ \citep[][]{Tang2026_spurs_NC, Marques-Chaves2024}. SPURS-GN-2 presents an inverted \ion{C}{iii}] doublet ratio with respect to COLA1, which suggest that significant carbon emission is produced in the dense gas ($\logten(\eden\rm / cm^{-3})\gtrsim 6.5$).  In turn, CEERS-1019 shows $\ion{C}{iii} 1906/1909\approx 1$; this observed ratio would be compatible with a single-zone density of $\approx10^4$~\unit{cm^{-3}}. However in a two-zone scenario as proposed for COLA1, this would be degenerate with a low density zone producing most of the \ion{C}{iii}] emission with some contribution from the high density zone (the same that dominates the \ion{N}{iv}] emission). A similar case is that of GN-z11, presenting a similar spectrum as CEERS-1019, and line ratio $\ion{C}{iii}]\sim 1$ \citep[][]{Chen2026_gnz11}.

Another noteworthy difference between COLA1, CEERS-1019, and SPURS-GN-2 is the \ion{C}{iv} profile. SPURS-GN-2 shows absorption imprinted on broad \ion{C}{iv}, suggesting that the UV is dominated by the central engine rather than a stellar population (Fig.~\ref{fig:lines_spurs_comp}). COLA1 notably shows enhanced emission near systemic velocities in \ion{C}{iv}. This is in line with COLA1 having a very high \fesclyc{}, as clear channels in the ISM facilitate the escape of the resonant \ion{C}{iv} photons (\citealt{Naidu2022}; see further discussion in Sect.~\ref{sec:reionization}).

\subsection{Ionizing properties of COLA1}\label{sec:ionizing_properties}

The new NIRSpec UV spectrum confirms the hypothesis that COLA1 is a strong producer and leaker of ionizing photons. In the first place, there is a very luminous \lya{} line as observed by previous works \citep[][]{Hu2016, Matthee2018} with $\rm EW_0 = 400^{+40}_{-30}$~\AA{}. In Fig.~\ref{fig:lya_orders} we show the \lya{} line profile from JWST/NIRSpec data, where the increased spectral resolution of the 2nd and 3rd order spectrum reveals the double-peak shape. We estimate the \lya{} escape fraction within the NIRSpec aperture comparing to the narrow \Halpha{} flux, assuming Case B recombination ($\lya/\Halpha \approx 8.6$) and negligible dust attenuation (Sect.~\ref{sec:balmer_lines_modeling}). This calculation assumes that the \lya{} is solely powered by the same ionizing source that powers the narrow component of \Halpha{}, which could be a combination of stars and any ionizing output from the central engine of the LRD. We obtain an escape fraction of $\fesclya= 0.68 \pm 0.03$. As discussed by \citet{Torralba2024}, the high \lya{} escape fraction, and narrow peak separation is suggestive of a high escape fraction of ionizing photons \citep[][]{Izotov2018, Maji2022, Kimm2022, Naidu2022}. We estimate the escape fraction using the models of \citet{Mascia2025}, which employ Cox models to infer \fesclyc{} using as inputs: $E(B-V)$, \MUV{}, stellar mass and $\logten(\Hbeta+[\ion{O}{iii}])$. With this model we obtain $\fesclyc = 0.67 \pm 0.04$.

We estimate the ionizing photon production efficiency as
\begin{equation}
\xi_{\rm ion} = L_\Halpha \cdot c_\Halpha{}^{-1} \cdot {L^\lambda_{\rm UV}}^{-1} \cdot (1 - \fesclyc)^{-1} \,,
\end{equation}
using a recombination coefficient $c_\Halpha{} = 1.128\times10^{-12}$~\unit{erg^{-1}} for $\Hden = 10^3$~\unit{cm^{-3}} and $T_{\rm eff} = 2\times 10^4$~K from PyNeb \citep[][]{Luridiana2015}. We use only the narrow \Halpha{} flux in this calculation, implicitly assuming that the UV is only of stellar origin, and that the narrow \Halpha{} is powered by the same stars. We obtain $\logten(\xi_{\rm ion,0}/\rm Hz\,erg^{-1})= 25.52 \pm 0.02$; where $\xi_{\rm ion, 0} = \xi_{ion} \cdot (1 - \fesclyc)$. If we use the previously quoted number of $\fesclyc=0.67$, the intrinsic ionizing production efficiency estimated is $\logten(\xiion/\rm Hz\,erg^{-1}) \approx 26$. Our fiducial SSP population including VMS (Sect.~\ref{sec:ficus_analysis}) yields an intrinsic $\logten(\xi_{\rm ion}/\rm Hz\,erg^{-1})= 25.77$, which would imply a somewhat lower escape fraction of $\fesclyc \approx 0.44$, albeit still high compared to confirmed leakers at low-$z$ \citep[e.g.,][]{Flury2022}. This discrepancy could also be explained by some amount of ionizing photons escaping from the LRD engine.

\subsection{Impact on cosmic reionization}\label{sec:reionization}

\begin{figure}
    \centering
    \includegraphics[width=\linewidth]{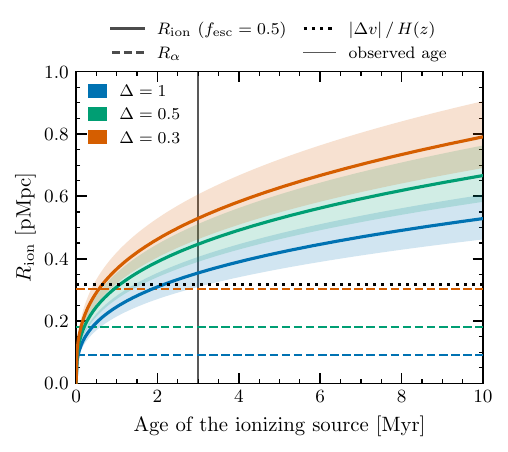}
    \caption{Growth of the ionized region around COLA1 since the start of the first starburst. We assume a isotropic ionization as a Str\"omgren sphere; further details of the model are given in Sect.~\ref{sec:reionization}. We show the required size of the proximity zone required to observe significant \lya{} flux ($>10\%$ transmission) at $\Delta v < -250$~\unit{km.s^{-1}}. The size of the proximity zone (neutral density $x_\text{H\,{\sc i}}<10^{-6}$) for each IGM density is shown as a dashed black line.}
    \label{fig:bubble_size}
\end{figure}

We investigate whether COLA1 is capable of carving the ionized region needed in order to explain the observed double-peaked \lya{} profile. We explore the growth of a spherical ionized region powered by a central source as a Str\"omgren sphere \citep[][]{Stromgren1939}. We assume $\logten(\xi_{\rm ion, 0}/\rm Hz\,erg^{-1}) = 25.52$ (Sect.~\ref{sec:ionizing_properties}) to compute the escaping ionizing photon rate as 
\begin{equation}
Q_{\rm esc}({\rm H}) = \xi_{\rm ion, 0} \cdot \frac{\fesclyc}{1 - \fesclyc} \cdot L_\nu (\text{1500\AA}) \,,
\end{equation}
defined as the number of photons produced by the source that are able to escape to the IGM and power an ionized bubble.
We assume a mean IGM density of $\langle n \rangle =1.88 \times 10^{-7} \cdot (1 + z)^3$~\unit{cm^{-3}} \citep[][]{Mason2020}. In Fig.~\ref{fig:bubble_size} we show the evolution of the ionized region for different values of \fesclyc{}, and for three fractions of the mean IGM density ($\Delta\equiv n_{\rm H}^{\rm IGM} / \langle n_{\rm H}^{\rm IGM} \rangle$), and assuming a clumping factor $C=3$ \citep[e.g.,][]{Finlator2012}. We compare these growth curves with the expected ionized region size in the line of sight that is needed to achieve sufficient transmission of the blue \lya{} peak. The significant \lya{} transmission down to $-250$~\unit{km.s^{-1}} implies the existence of a proximity zone of at least $\sim 0.32$~pMpc \citep[][]{Matthee2018}. We compute the size of the proximity zone $R_\alpha$ of our ionization model \citep[][]{Bolton2007}. The size of the proximity zone depends on the IGM density, roughly as $\propto \Delta^{-1}$. As shown previously by \citet{Mason2020}, there needs to be an underdensity of $\Delta\lesssim 0.5$ in order for a source like COLA1 to sustain a proximity zone of $>0.32$~pMpc, but the total ionized region must be larger if the effects of a \lya{} damping wing by the neutral IGM is taken into account. In particular, \citet{Mason2020} derived a size of $R_\alpha \approx 0.7$~Mpc, assuming that the nearest neutral patch has $\Delta = 1$. However, the $R_\alpha$ requirement can be relaxed towards smaller ionized regions if the density of the neutral IGM is lower. A complete IGM-ionization study is out of the scope of this work. To show consistency with our results, in Fig.~\ref{fig:bubble_size} we show that COLA1 is able to ionize past the required proximity zone of 0.32~Mpc for an underdense sightline of $\Delta\gtrsim 0.5$ in a timescale much shorter than the age of the stellar population ($t\lesssim 2$~Myr). However, the requirement of significant ionization inside the bubble to sustain a \lya{} transparent zone implies even stronger underdensities. As shown in Fig.~\ref{fig:bubble_size}, an underdensity as low as $\Delta = 0.3$ is needed so COLA1 can sustain a \lya{} transparent proximity zone to match the observed blue peak transmission of $\sim 250$~\unit{km.s^{-1}}. Such strong constraints would require a very rare underdense line of sight, for example, placing COLA1 at the edge of a cosmic void as discussed by \citet{Park2026_cola1}.

Interestingly, the galaxies CEERS-1019 and GN-z11, which have very similar UV spectra to COLA1 (see discussion in Sect.~\ref{sec:compare_spurs}) are also strong \lya{} emitters at $z>7$. \citet{Chen2026_gnz11} have discussed that GN-z11 must reside in an ionized bubble of $\sim 0.5$~pMpc, and \citet{Whitler2026} found evidence for a $\sim 0.5$--1~pMpc ionized bubble at $z=8.7$ in an overdensity containing CEERS-1019. Moreover, the \lya{} observed in some high-$z$ LRDs suggest that some of these sources also trace large ionized regions \citep[][]{Tang2026_clumpy, Tang2026_spurs_NC}. All these results suggest a connection between the LRD and N-enhanced galaxy phenomenon and the presence of large ionized IGM regions.
It is possible that a previous burst of ionizing photons powered the proximity zone before turning off \citep[][]{Meyer2026_cola1}, which should have happened in the last $\sim 0.1$--0.7~Myr (for $\Delta=0.5$ to 0.2), before the IGM naturally recombines \citep[][]{Park2026_cola1}. This could suggest that there was a much more luminous episode of LRD or AGN activity in the past, or that there was recent even stronger burst of star formation. The latter case could be represented by the extremely luminous compact galaxies recently reported by \citet{Yang2026}, whose UV spectrum is dominated by the light of VMSs and presents very weak nebular emission lines pointing towards a very high \fesclyc{} \citep[see also][]{Marques-Chaves2026_fesc, Wu2026}. Given the high number densities of LRDs at $z>6$ \citep[e.g.,][]{Kocevski2025, Zhuang2026}, if an early COLA1-like phase turns out to be common around most LRDs, these objects have the potential to have played a critical role in local reionization through short starburst episodes with high ionizing output powering Mpc-scale bubbles \citep[e.g.,][]{Naidu2020, Matthee2022}.

\section{Summary}\label{sec:summary}

The spectrum of the galaxy COLA1 presents a striking combination: a UV spectrum dominated by very massive stars, a red LRD-like component in the optical, and extremely high nitrogen enhancement. Below we summarize our main findings.

\begin{enumerate}
    \item We fit the UV spectrum of COLA1 with stellar population synthesis models, favoring those containing VMS (limited to masses up to $300$~\Msun{}). We find a best-fit stellar population with mean age $2.43\pm0.75$~Myr, and stellar mass $\approx10^8$~\Msun{} (for an IMF with slope $\alpha=2.35$). In particular, the VMSs are needed to reproduce the P~Cygni-like profiles of \ion{N}{v} $\lambda\lambda 1239,1243$, \ion{Si}{iv}] $\lambda\lambda 1394,1403$, and \ion{C}{iv} $\lambda\lambda 1548,1551$, and the broad component of \ion{He}{ii} $\lambda 1640$ (Sect.~\ref{sec:ficus_analysis}).
    
    \item While the rest-UV spectrum has a fairly blue slope ($\beta_{\rm UV} = -2.5$), the rest-optical is significantly redder ($\beta_{\rm opt}=-1.1$), shows strong \ion{He}{i}, and broad \Hbeta{} and \Halpha{} with a narrow Balmer absorption. The narrow components of the Balmer lines follow Case B recombination ratios with little-to-no dust attenuation, while the broad components present high decrements ($\Halpha/\Hbeta\sim9$) that could be explained by attenuation by nuclear dust, but also by RT effects in a dense wind akin an LRD envelope (Sect.~\ref{sec:balmer_lines_modeling}).
    
    \item We measured the nitrogen and carbon abundance of COLA1 using a single-zone density model against line emissivities from PyNeb, obtaining a supersolar $\logNO = -0.07\pm0.05$, and sub-solar $\logCO = -0.64\pm10$. The high value of $\logNC = +0.58\pm10$ requires exotic enrichment channels, such as VMS winds, TDEs, or the yields of SMSs. Moreover, decomposing the spectrum with a two-zone model using a Cloudy grid, we find evidence for the presence of a dense zone ($\eden\approx 10^{7.5}$~\unit{cm^{-3}}) with extreme $\logNC>1.3$ (2$\sigma$), and a more diffuse zone with $\logNC< +0.02$. Such abundances can only be explained by the most extreme enrichment channels such as $>10^4$~\Msun{} SMSs (Sects.~\ref{sec:cloudy_line_modeling}, \ref{sec:abundances}, and \ref{sec:NC_yields_model}).

    \item We argue that the UV-to-optical spectrum of COLA1 can be well described by the combination of a dense stellar cluster of young massive stars, including VMSs ($>100$~\Msun{}), dominating the UV light; and a red LRD component rising towards the rest-optical wavelength range. The lives of the VMSs that dominate the UV are predicted to be short ($\sim 1$--2~Myr), therefore the UV of this system will fade by 1--2~mag in timescales of $\sim5$~Myr (depending on the IMF), then COLA1 would be observed as a typical V-shaped LRD (Sects.~\ref{sec:LRD_plus_vms} and \ref{sec:vms_evolution}).

    \item We discuss possible mechanisms leading to the single- and two-zone N/C abundances observed in COLA1. The winds of rapidly rotating VMS could account for the measured single-zone abundance only under very optimistic conditions. Producing such high $\logNC\approx 0.6$ with stellar yields would require models of $\gtrsim 1000$~\Msun{} Pop\,III stars, or SMSs with $10^\text{4--5}$~\Msun{}. In particular, the inferred dense zone abundance of $\logNC>1.3$ (2$\sigma$) requires enrichment from the yiels of SMSs, or TDEs of massive stars formed in a dense star cluster (Sects.~\ref{sec:abundances}).

    \item We speculate that runaway collisions in a dense star cluster could have planted a massive black hole seed, that manifests in the observed LRD-like spectrum through dense-gas enshrouded accretion into a newly formed IMBH; ultimately leading to the early stages of the formation of a SMBH. The same process may have formed the VMSs observed in the UV as a by-product (Sect.~\ref{sec:interpretation}).

    \item COLA1 also shows extraordinary ionizing properties, a high inferred escape fraction ($\fesclyc\approx 0.67$), and lives inside a $>0.3$~Mpc highly ionized bubble. We argue that if COLA1 represents a common phase in the LRD evolution, given the elevated number densities at $z>6$, this population may have a crucial impact on cosmic reionization. The light of the observed VMSs, but possibly also past intense starbursts may have been responsible for powering COLA1's ionized bubble (Sects.~\ref{sec:compare_spurs}, \ref{sec:ionizing_properties}, and \ref{sec:reionization}).
\end{enumerate}

% JWST is forcing us to revise our understanding of early stellar populations and the birth of the present-day supermassive black holes.
JWST is revealing the remarkable, unprecedented properties of early stellar populations. Here, we show that these stellar populations are intimately linked to the perplexing BH populations at high-$z$ such as LRDs, and may be critical to unraveling the seeding and birth of present-day massive BHs.
Recent studies are revealing that populations of VMS are common in the most luminous high-$z$ galaxies \citep[][]{Upadhyaya2024, MarquesChaves2026_vms, Nakane2026_gnz11, Chen2026_gnz11, Yang2026}. With COLA1, we may be witnessing how the same dense environment harbouring a population of VMSs is the melting pot in which a supermassive star can form and collapse, seeding the massive black holes of the nearby Universe.

\section*{Acknowledgments}

AT thanks Debasish Dutta, Ruggero Valli, Sunmyon Chon, and Zolt\'an Haiman for useful conversations about stellar winds of massive stars, stellar evolution, supermassive star formation and massive black hole seeding.
We acknowledge funding by the European Union (ERC, AGENTS,  101076224).
This work is based on observations made with the NASA/ESA/CSA James Webb Space Telescope. The data were obtained from the Mikulski Archive for Space Telescopes at the Space Telescope Science Institute, which is operated by the Association of Universities for Research in Astronomy, Inc., under NASA contract NAS 5-03127 for JWST.
These observations are associated with programs \#7722, \#3767, and \#1933.
This paper makes use of the following ALMA data: ADS/JAO.ALMA\#2019.1.00741.S ALMA is a partnership of ESO (representing its member states), NSF (USA) and NINS (Japan), together with NRC (Canada), NSTC and ASIAA (Taiwan), and KASI (Republic of Korea), in cooperation with the Republic of Chile. The Joint ALMA Observatory is operated by ESO, AUI/NRAO and NAOJ.
Some of the data products presented herein were retrieved from the Dawn JWST Archive (DJA). DJA is an initiative of the Cosmic Dawn Center (DAWN), which is funded by the Danish National Research Foundation under grant DNRF140.
Generative AI (Anthropic Claude Opus 5.5) assisted with coding and literature searching. The authors reviewed and approved all scientific content and take full responsibility for the manuscript. The text was entirely written, edited, and reviewed by humans.

\bibliographystyle{bibtex/aa.bst}
\bibliography{bibliography}

\begin{appendix}

\section{Emission line additional flux measurements}\label{app:emline_fluxes}

\begin{figure*}
    \centering
    \includegraphics[width=\linewidth]{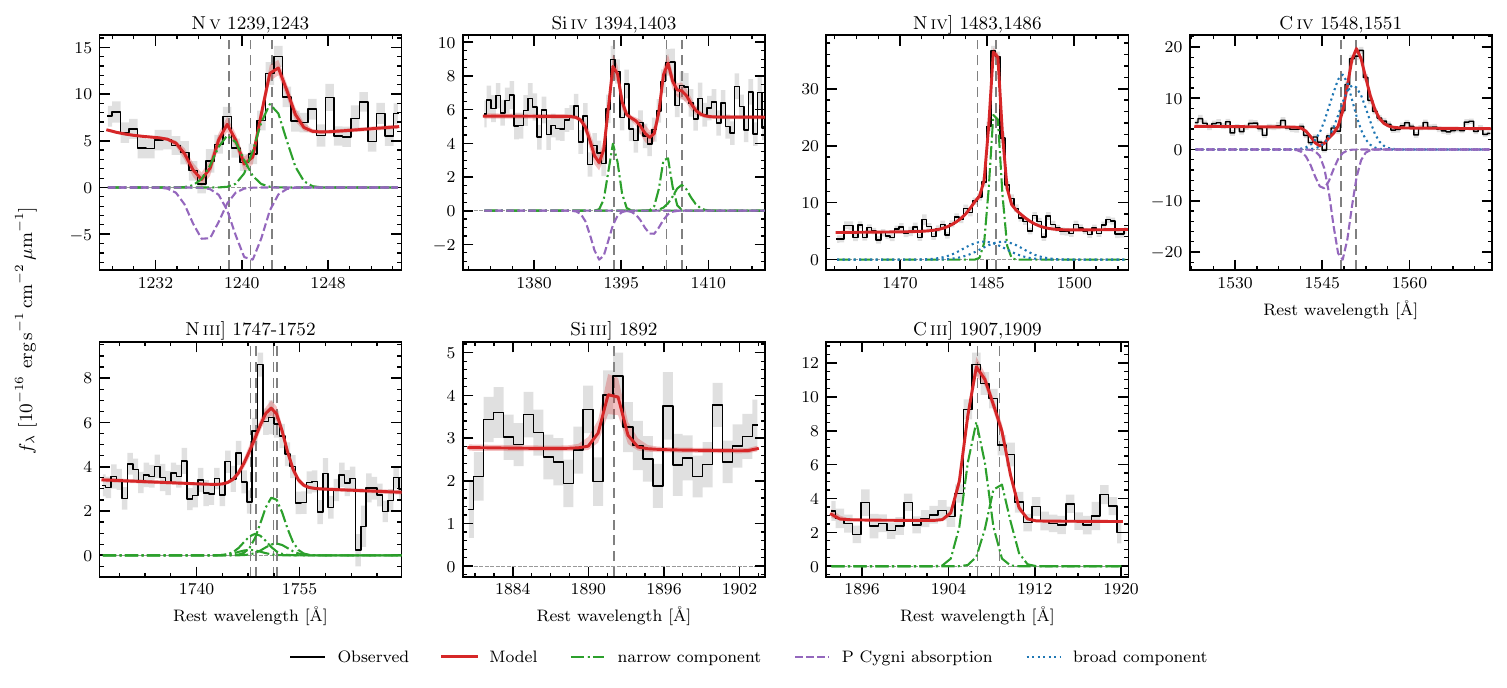}
    \caption{Fit to the rest-UV emission lines. As detailed in Sect.~\ref{sec:emline_measurements}, we use Gaussians for all the emission lines, adding absorbers and broad components when relevant.}
    \label{fig:all_uv_lines}
\end{figure*}

\begin{figure*}
    \centering
    \includegraphics[width=\linewidth]{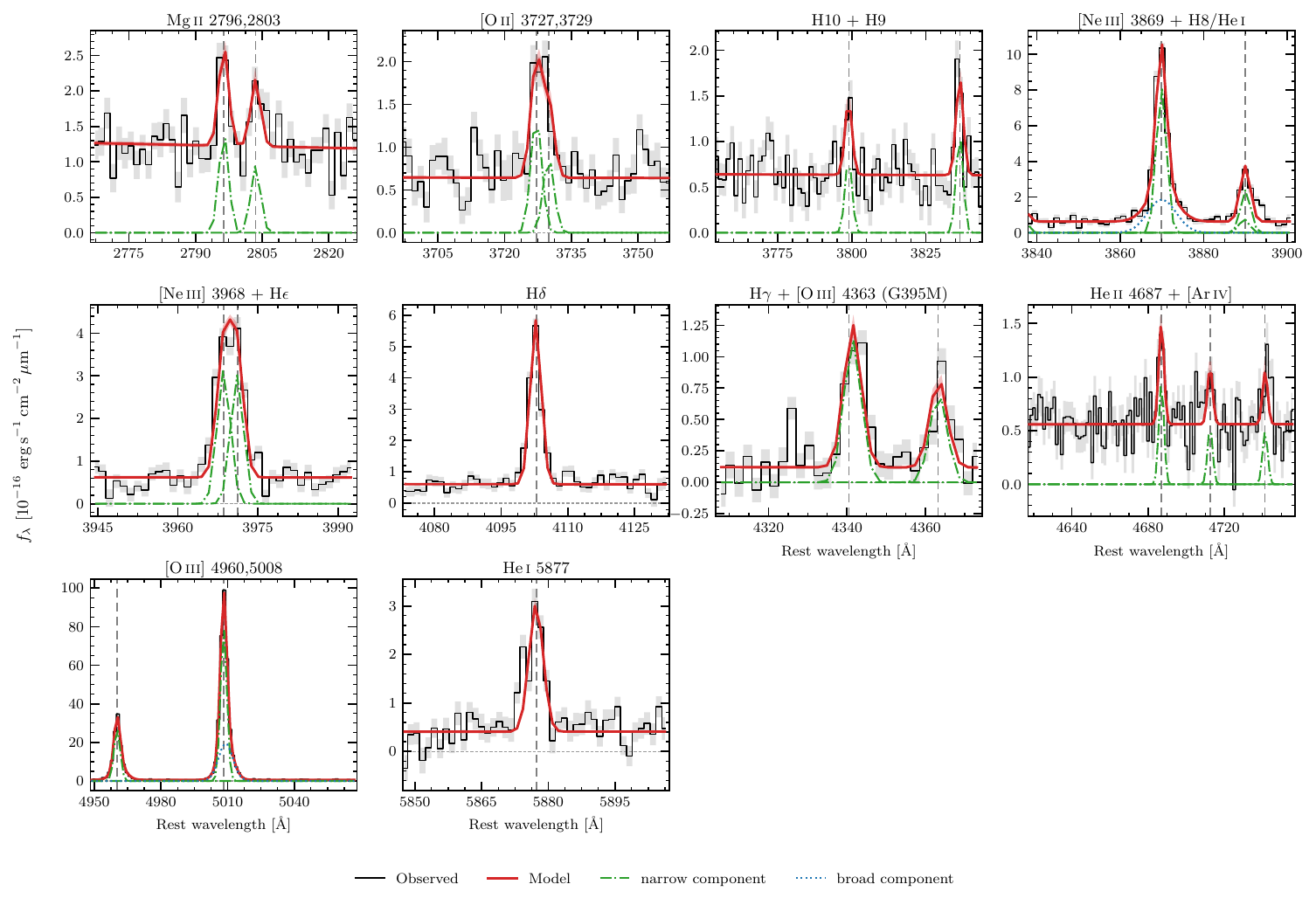}
    \caption{Same as Fig.~\ref{fig:all_uv_lines}, for the rest-optical lines. The Balmer lines with detected broad components (\Halpha{}, \Hbeta{}) are fitted separately as described in Sect.~\ref{sec:balmer_lines_modeling}.}
    \label{fig:all_optical_lines}
\end{figure*}

In Figs.~\ref{fig:all_uv_lines} and \ref{fig:all_optical_lines} we show the UV and optical emission line fits, respectively. The models and procedure are described in Sect.~\ref{sec:emline_measurements}. We report the flux, rest-frame EW and Gaussian FWHM of all the fitted components in Tables~\ref{tab:emlines_uv} and \ref{tab:emlines_opt}.

\begin{table}
\centering
\caption{Ultraviolet emission line measurements for COLA1 ($z=6.590537 \pm 0.000015$).}
\label{tab:emlines_uv}
\begin{tabular}{lccccc}
\hline
Line & $\lambda_\mathrm{rest}$ [\AA] & Flux & EW$_0$ & FWHM \\
\hline
{\lya{}} &  1215.67 & $706^{+5}_{-6}$ & $400^{+40}_{-30}$ & $280^{+30}_{-30}$ \\
{\ion{N}{v}} &  1238.82 & $7^{+10}_{-4}$ & $2^{+3}_{-1}$ & $660^{+150}_{-120}$ \\
{\ion{N}{v} (abs)} &  1238.82 & $-11^{+3}_{-5}$ & $-3.2^{+0.7}_{-1.4}$ & $580^{+80}_{-160}$ \\
{\ion{N}{v}} &  1242.80 & $22^{+4}_{-3}$ & $5.8^{+1.1}_{-1.0}$ & $660^{+150}_{-120}$ \\
{\ion{N}{v} (abs)} &  1242.80 & $-15^{+4}_{-6}$ & $-4.2^{+1.1}_{-1.6}$ & $600^{+80}_{-120}$ \\
{\ion{Si}{iv}} &  1393.76 & $6^{+3}_{-1}$ & $1.5^{+0.7}_{-0.3}$ & $350^{+130}_{-130}$ \\
{\ion{Si}{iv} (abs)} &  1393.76 & $-7.9^{+1.7}_{-1.9}$ & $-1.9^{+0.4}_{-0.4}$ & $640^{+50}_{-90}$ \\
{\ion{Si}{iv}} &  1402.77 & $5^{+2}_{-2}$ & $1.3^{+0.6}_{-0.4}$ & $350^{+130}_{-130}$ \\
{\ion{Si}{iv} (abs)} &  1402.77 & $-3.6^{+1.3}_{-2.0}$ & $-0.9^{+0.3}_{-0.4}$ & $570^{+100}_{-180}$ \\
{[\ion{O}{iv}]} &  1405.46 & $4.2^{+1.9}_{-1.6}$ & $1.0^{+0.4}_{-0.4}$ & $700^{+200}_{-300}$ \\
{[\ion{N}{iv}]} &  1483.32 & $1.4^{+1.3}_{-1.0}$ & $0.4^{+0.3}_{-0.3}$ & $360^{+30}_{-30}$ \\
{[\ion{N}{iv}] (broad)} &  1483.32 & $24^{+16}_{-16}$ & $6^{+4}_{-4}$ & $1800^{+200}_{-200}$ \\
{\ion{N}{iv}]} &  1486.50 & $48^{+3}_{-4}$ & $12.6^{+0.9}_{-0.9}$ & $360^{+30}_{-30}$ \\
{\ion{N}{iv}] (broad)} &  1486.50 & $20^{+16}_{-16}$ & $5^{+4}_{-4}$ & $1800^{+200}_{-200}$ \\
{\ion{C}{iv}} &  1548.19 & $0.6^{+1.0}_{-0.5}$ & $0.2^{+0.3}_{-0.1}$ & $250^{+40}_{-90}$ \\
{\ion{C}{iv} (abs)} &  1548.19 & $-23^{+6}_{-9}$ & $-7^{+2}_{-3}$ & $680^{+10}_{-20}$ \\
{\ion{C}{iv} (broad)} &  1548.19 & $70^{+60}_{-70}$ & $20^{+20}_{-20}$ & $1040^{+50}_{-30}$ \\
{\ion{C}{iv}} &  1550.77 & $1.1^{+1.6}_{-0.8}$ & $0.3^{+0.5}_{-0.2}$ & $250^{+40}_{-90}$ \\
{\ion{C}{iv} (abs)} &  1550.77 & $-64^{+16}_{-20}$ & $-20^{+5}_{-6}$ & $680^{+10}_{-20}$ \\
{\ion{C}{iv} (broad)} &  1550.77 & $60^{+50}_{-50}$ & $17^{+16}_{-15}$ & $1040^{+50}_{-30}$ \\
{\ion{He}{ii}} &  1640.42 & $3.7^{+1.2}_{-1.2}$ & $1.3^{+0.4}_{-0.4}$ & $244^{+22}_{-24}$ \\
{\ion{He}{ii} (broad)} &  1640.42 & $7.4^{+1.5}_{-1.7}$ & $2.5^{+0.5}_{-0.6}$ & $970^{+280}_{-200}$ \\
{\ion{O}{iii}]} &  1660.81 & $4.8^{+0.7}_{-0.6}$ & $1.7^{+0.2}_{-0.2}$ & $170^{+30}_{-40}$ \\
{\ion{O}{iii}]} &  1666.15 & $12.7^{+0.7}_{-0.7}$ & $4.5^{+0.3}_{-0.3}$ & $170^{+30}_{-40}$ \\
{\ion{N}{iii}]} &  1747.85 & $0.8^{+1.1}_{-0.6}$ & $0.3^{+0.5}_{-0.3}$ & $650^{+80}_{-70}$ \\
{\ion{N}{iii}]} &  1748.65 & $3.3^{+1.5}_{-1.5}$ & $1.4^{+0.6}_{-0.6}$ & $650^{+80}_{-70}$ \\
{\ion{N}{iii}]} &  1751.22 & $9^{+2}_{-3}$ & $3.6^{+0.8}_{-1.1}$ & $650^{+80}_{-70}$ \\
{\ion{N}{iii}]} &  1751.66 & $2^{+2}_{-1}$ & $0.7^{+1.1}_{-0.5}$ & $650^{+80}_{-70}$ \\
{\ion{Si}{iii}]} &  1892.03 & $2.2^{+0.7}_{-0.8}$ & $1.1^{+0.4}_{-0.4}$ & $200^{+200}_{-100}$ \\
{[\ion{C}{iii}]} &  1906.68 & $16.1^{+0.9}_{-1.1}$ & $7.9^{+0.5}_{-0.6}$ & $310^{+40}_{-40}$ \\
{\ion{C}{iii}]} &  1908.73 & $9.4^{+1.0}_{-0.9}$ & $4.6^{+0.5}_{-0.5}$ & $310^{+40}_{-40}$ \\
{[\ion{Ne}{iv}]} &  2424.26 & $<1.7$ & $<1.4$ & $133^{+4}_{-4}$ \\
{\ion{Mg}{ii}]} &  2796.35 & $3.0^{+0.4}_{-0.4}$ & $2.8^{+0.4}_{-0.3}$ & $132^{+4}_{-4}$ \\
{\ion{Mg}{ii}]} &  2803.53 & $2.0^{+0.3}_{-0.3}$ & $1.8^{+0.4}_{-0.2}$ & $132^{+4}_{-4}$ \\
{[\ion{Fe}{iv}]} &  2830.19 & $1.2^{+0.3}_{-0.4}$ & $1.0^{+0.4}_{-0.3}$ & $132^{+4}_{-4}$ \\
{[\ion{Fe}{iv}]} &  2836.57 & $2.6^{+0.4}_{-0.4}$ & $2.4^{+0.5}_{-0.3}$ & $132^{+4}_{-4}$ \\
{[\ion{Ar}{iv}]} &  2854.51 & $0.6^{+0.3}_{-0.3}$ & $0.4^{+0.3}_{-0.2}$ & $132^{+4}_{-4}$ \\
{[\ion{Ar}{iv}]} &  2869.05 & $0.13^{+0.16}_{-0.10}$ & $0.04^{+0.11}_{-0.03}$ & $132^{+4}_{-4}$ \\
{\ion{He}{i}} &  3188.67 & $1.6^{+0.3}_{-0.3}$ & $1.8^{+0.4}_{-0.2}$ & $132^{+4}_{-4}$ \\
{[\ion{Ne}{v}]} &  3426.86 & $<0.8$ & $<1.2$ & $133^{+4}_{-4}$ \\
\hline
\end{tabular}
\tablefoot{
Fluxes are in units of $10^{-19}$\,erg\,s$^{-1}$\,cm$^{-2}$, rest-frame EW in \AA, and FWHM in km\,s$^{-1}$.
Upper limits are given for a 3$\sigma$ confidence level.
}
\end{table}

\begin{table}
\centering
\caption{Optical emission line measurements for COLA1 ($z=6.590537 \pm 0.000015$).}
\label{tab:emlines_opt}
\begin{tabular}{lccccc}
\hline
Line & $\lambda_\mathrm{rest}$ [\AA] & Flux & EW$_0$ & FWHM \\
\hline
{[\ion{O}{ii}]} &  3727.09 & $3.2^{+0.4}_{-0.4}$ & $5.7^{+0.8}_{-0.5}$ & $132^{+4}_{-4}$ \\
{[\ion{O}{ii}]} &  3729.88 & $2.0^{+0.4}_{-0.4}$ & $3.4^{+0.7}_{-0.5}$ & $132^{+4}_{-4}$ \\
{H(10-2)} &  3798.98 & $1.9^{+0.3}_{-0.3}$ & $3.1^{+0.8}_{-0.5}$ & $122^{+6}_{-6}$ \\
{H(9-2)} &  3836.48 & $2.5^{+0.3}_{-0.4}$ & $4.4^{+0.7}_{-0.4}$ & $122^{+6}_{-6}$ \\
{[\ion{Ne}{iii}]} &  3869.86 & $19.1^{+1.0}_{-1.0}$ & $38^{+2}_{-2}$ & $132^{+4}_{-4}$ \\
{[\ion{Ne}{iii}] (broad)} &  3869.86 & $12.3^{+1.1}_{-1.1}$ & $23^{+2}_{-2}$ & $590^{+70}_{-60}$ \\
{\ion{He}{i}} &  3889.75 & $1.8^{+1.7}_{-1.3}$ & $1^{+3}_{-1}$ & $132^{+4}_{-4}$ \\
{H(8-2)} &  3890.17 & $5.4^{+1.4}_{-1.7}$ & $8^{+4}_{-2}$ & $122^{+6}_{-6}$ \\
{[\ion{Ne}{iii}]} &  3968.59 & $7.5^{+0.4}_{-0.4}$ & $15.1^{+0.9}_{-0.6}$ & $132^{+4}_{-4}$ \\
{H$\varepsilon$} &  3971.20 & $7.0^{+0.5}_{-0.4}$ & $14.0^{+0.9}_{-0.8}$ & $122^{+6}_{-6}$ \\
{H$\delta$} &  4102.90 & $12.3^{+0.4}_{-0.4}$ & $25.8^{+1.1}_{-0.7}$ & $122^{+6}_{-6}$ \\
{H$\gamma$} & 4341.68 & $21.4^{+1.6}_{-1.6}$ & $47^{+6}_{-4}$  & $122^{+6}_{-6}$\\
{[\ion{O}{iii}]} & 4363.21 & $13.2^{+1.6}_{-1.8}$ & $28^{+5}_{-3}$ & $121^{+3}_{-4}$\\
{\ion{He}{ii}} &  4687.02 & $2.5^{+0.3}_{-0.3}$ & $4.9^{+0.8}_{-0.5}$ & $132^{+4}_{-4}$ \\
{[\ion{Ar}{iv}]} &  4712.58 & $1.4^{+0.3}_{-0.3}$ & $2.5^{+0.8}_{-0.4}$ & $132^{+4}_{-4}$ \\
{[\ion{Ar}{iv}]} &  4741.45 & $1.3^{+0.3}_{-0.3}$ & $2.3^{+0.8}_{-0.5}$ & $132^{+4}_{-4}$ \\
{\Hbeta} &  4862.69 & $36.8^{+1.0}_{-1.0}$ & $83^{+3}_{-2}$ & $122^{+6}_{-6}$ \\
{\Hbeta{}*} &  4862.69 & $9^{+4}_{-4}$ & $13^{+9}_{-5}$ & $2900^{+70}_{-110}$ \\
{\Hbeta{}*} &  4862.69 & $8.1^{+1.2}_{-1.1}$ & $16^{+3}_{-1}$ & $330^{+20}_{-20}$ \\
{\Hbeta{}*} &  4862.69 & $15^{+3}_{-3}$ & $29^{+7}_{-5}$ & $1460^{+100}_{-110}$ \\
{[\ion{O}{iii}]} &  4960.30 & $71.7^{+1.4}_{-1.1}$ & $163^{+5}_{-3}$ & $132^{+4}_{-4}$ \\
{[\ion{O}{iii}] (broad)} &  4960.30 & $40.5^{+1.1}_{-1.3}$ & $91^{+4}_{-2}$ & $431^{+7}_{-7}$ \\
{[\ion{O}{iii}]} &  5008.24 & $214^{+4}_{-3}$ & $484^{+17}_{-9}$ & $132^{+4}_{-4}$ \\
{[\ion{O}{iii}] (broad)} &  5008.24 & $120.7^{+3}_{-4}$ & $270^{+12}_{-7}$ & $431^{+7}_{-7}$ \\
{\ion{He}{i}} &  5877.25 & $8.0^{+0.6}_{-0.5}$ & $24^{+2}_{-1}$ & $132^{+4}_{-4}$ \\
{[\ion{N}{ii}]} &  6549.86 & $1.9^{+1.0}_{-1.0}$ & $2^{+2}_{-1}$ & $132^{+4}_{-4}$ \\
{\Halpha{}} &  6564.64 & $121^{+4}_{-5}$ & $286^{+20}_{-12}$ & $122^{+6}_{-6}$ \\
{\Halpha{}*} &  6564.64 & $64^{+6}_{-5}$ & $146^{+16}_{-9}$ & $330^{+20}_{-20}$ \\
{\Halpha{}*} &  6564.64 & $92^{+16}_{-15}$ & $190^{+40}_{-30}$ & $1460^{+100}_{-110}$ \\
{\Halpha{}*} &  6564.64 & $195^{+17}_{-18}$ & $430^{+60}_{-40}$ & $2900^{+70}_{-110}$ \\
{[\ion{N}{ii}]} &  6585.28 & $0.6^{+0.6}_{-0.4}$ & $0.5^{+1.0}_{-0.3}$ & $132^{+4}_{-4}$ \\
{\ion{He}{i}} &  6680.00 & $2.5^{+0.8}_{-0.7}$ & $5^{+2}_{-1}$ & $132^{+4}_{-4}$ \\
\hline
\end{tabular}
\tablefoot{
Fluxes are in units of $10^{-19}$\,erg\,s$^{-1}$\,cm$^{-2}$, rest-frame EW in \AA, and FWHM in km\,s$^{-1}$.
This components are modeled as broad Gaussians to capture the general shape of the lines, a dedicated modeling of \Halpha{} and \Hbeta{} in particular is detailed in Sect.~\ref{sec:balmer_lines_modeling}.
}
\end{table}

\section{\Halpha{} alternative fits}

\begin{figure*}
    \centering
    \includegraphics[width=\linewidth]{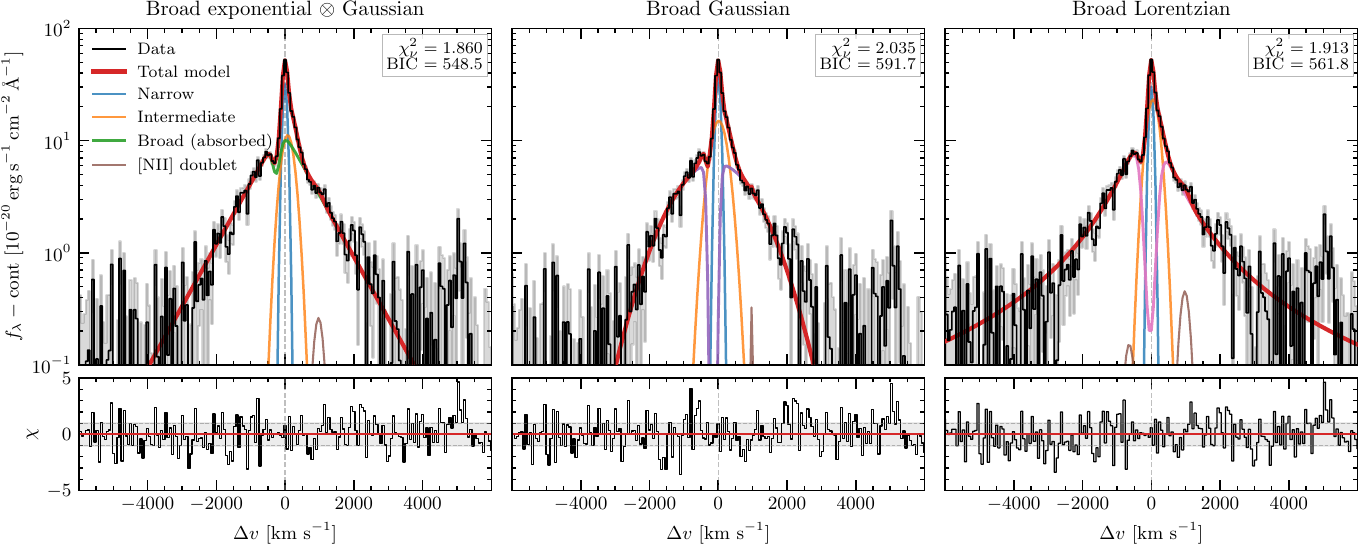}
    \caption{Best-fit models to the \Halpha{} line-profile using the models in \cite{Torralba2026_gn9771, Matthee2026}, tailored for Balmer lines of LRDs. From left to right, we use a symmetric exponential, Gaussian, and Lorentzian profile to model the broad component of the line, respectively. We find that an exponential provides the best fit, with marginal improvement with respect to a Lorentzian as a second-best option ($\rm\Delta BIC = +13$).}
    \label{fig:Ha_fit_LRD_model}
\end{figure*}

In Fig.~\ref{fig:Ha_fit_LRD_model} we show the fit to the \Halpha{} emission line profile of COLA1. The model is described in Sect.\ref{sec:balmer_lines_modeling}, but see also \citet{Torralba2026_gn9771} and \citet{Matthee2026}.

\section{Two-zone abundance fits}\label{app:abundances}

\begin{figure*}
    \centering
    \includegraphics[width=0.8\linewidth]{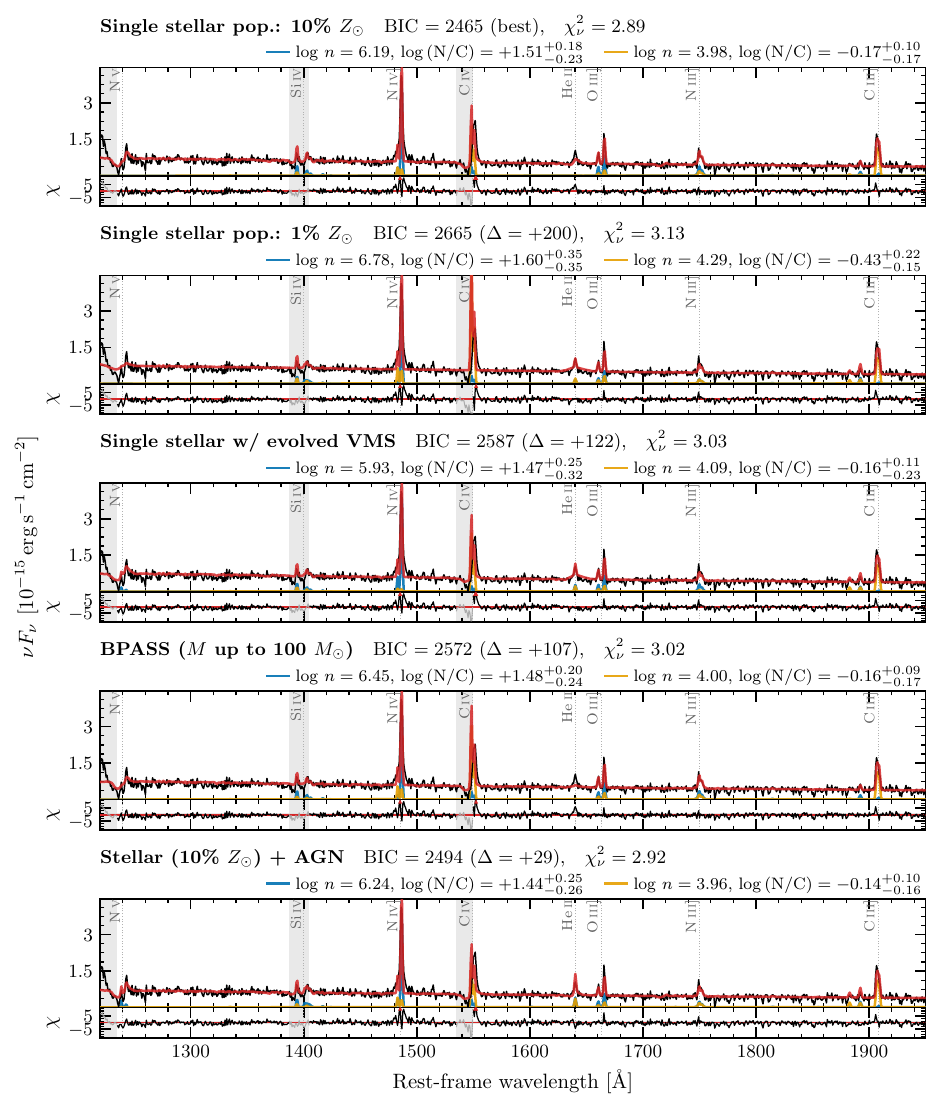}
    \caption{Two-zone Cloudy model fits to the UV spectrum. The models are described in Sect.~\ref{sec:cloudy_line_modeling}. Different panels show different intrinsic SEDs for the Cloudy runs, using different assumptions of the metallicity of the stellar population, using VMS or not, including evolved VMS, and including an AGN component (see labels in the legend).}
    \label{fig:cloudy_emlines}
\end{figure*}

Fig.~\ref{fig:cloudy_emlines} shows the two-zone decomposition using Cloudy models to the UV spectrum of COLA1. This analysis is detailed in Sect.~\ref{sec:cloudy_line_modeling}.

\section{NIRSpec G140M \lya{} spectrum}

\begin{figure}
    \centering
    \includegraphics[width=\linewidth]{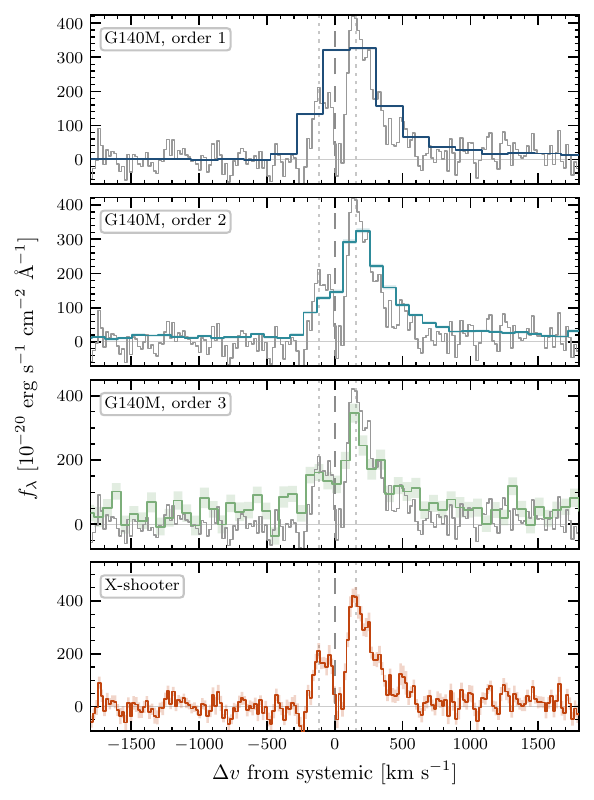}
    \caption{We show the 1st, 2nd, and 3rd order spectrum of \lya{} in the G140M spectrum of COLA1. Increasing orders show higher resolution, revealing the double peaked profile of the line observed by VLT/X-Shooter data \citep[][]{Matthee2018, Torralba2024}. The subsequent orders are normalized in the range $-300<\Delta v < 400$~\unit{km.s^{-1}} to match the flux of the 1st one to account for uncertain high-order flux-loss. We normalize the X-Shooter spectrum in the same way to account for the differences in slit-losses of both spectra.}
    \label{fig:lya_orders}
\end{figure}

In Fig.~\ref{fig:lya_orders} we show the NIRSpec \lya{} profile of COLA1. Due to the redshift and the choice of the grating/filter combination G140M/F070LP, we detect the \lya{} line in the first-, second- and third-order spectral traces. The line profile is totally unresolved in the first order, but has increased resolution in subsequent orders as shown in Fig.~\ref{fig:lya_orders}. We compare with the VLT/X-Shooter $R\approx 4000$ spectrum, that clearly shows the double-peak \lya{} \citep[][]{Matthee2018, Torralba2024}.

\end{appendix}

\end{document}